\documentclass[a4paper,fleqn]{cas-sc}

\usepackage[numbers,sort&compress]{natbib}
\usepackage{float}

\def\tsc#1{\csdef{#1}{\textsc{\lowercase{#1}}\xspace}}
\tsc{WGM}
\tsc{QE}

\begin{document}
\let\WriteBookmarks\relax
\def\floatpagepagefraction{1}
\def\textpagefraction{.001}

\shorttitle{Nonlinear mechanics of Kresling origami}    

\shortauthors{Yue, \emph{et al.}}  

\title[mode = title] {Nonlinear Mechanics and Predictable Bifurcation of Multi-Cell Kresling Origami Chains}



%

\author[1]{Songlin Yue}
\affiliation[1]{organization={Institute for Infrastructure and Environment, School of Engineering, The University of Edinburgh},
            city={Edinburgh},
            postcode={EH9 3FG}, 
            country={UK}}
%



\credit{Conceptualization, methodology, developed the analysis, and wrote and edited the paper}

\author[1,2]{Leo {de Waal}}
\affiliation[2]{organization={Civil, Environmental and Geomatic Engineering, Faculty of Engineering Sciences, University College London},
            city={London},
            postcode={WC1E 7JE}, 
            country={UK}}



\credit{Conceptualization, methodology, supervision, and edited the paper}

\author[1]{David {Garcia Cava}}




\credit{Conceptualization, methodology, supervision, and edited the paper}
\author[1]{Marcelo A. Dias}
[type=editor, orcid=0000-0002-1668-0501]
\cormark[1]


\ead{marcelo.dias@ed.ac.uk}


\credit{Conceptualization, methodology, supervision, and edited the paper}

\cortext[1]{Corresponding author}



\begin{abstract}
Meta-structures that display axial-twist coupling can be achieved through the emerging kinematics in Kresling origami patterns. A central challenge in these structures is understanding their nonlinear mechanical behaviour, specifically their equilibrium branches and bifurcation diagrams. This involves identifying relationships between desired responses and the geometric variables that define the design space, including the Kresling polygon count, initial twist angle, height, radius, and crease lengths. As the number of constituent units increases in an $n$-layer chain, we track complex equilibrium branches extending into the post-critical regime under successive instabilities, including branch-point bifurcations and limit-point instabilities. This work begins by establishing the relationship between the geometric design variables and the response curves of the assembled chain by modelling the crease lines as axial-load-carrying elements. Subsequently, equilibrium branches and instabilities are systematically investigated via continuation and bifurcation analysis, beginning with the single-layer system and progressively extending to two- and three-layer configurations. Finally, a generalisation strategy is proposed to extend these findings to an $n$-layer Kresling chain. This strategy enables the predictive construction of equilibrium paths and the inverse design of multi-layer meta-structures, using prescribed critical points to control post-critical behaviour. It provides a foundation for the inverse design and optimisation of architected mechanical metamaterials with programmable responses.
\end{abstract}




\maketitle
\section{Introduction}
\label{sec:Introduction}

\begin{figure}
 \begin{center}
  \includegraphics[width=0.95\textwidth]{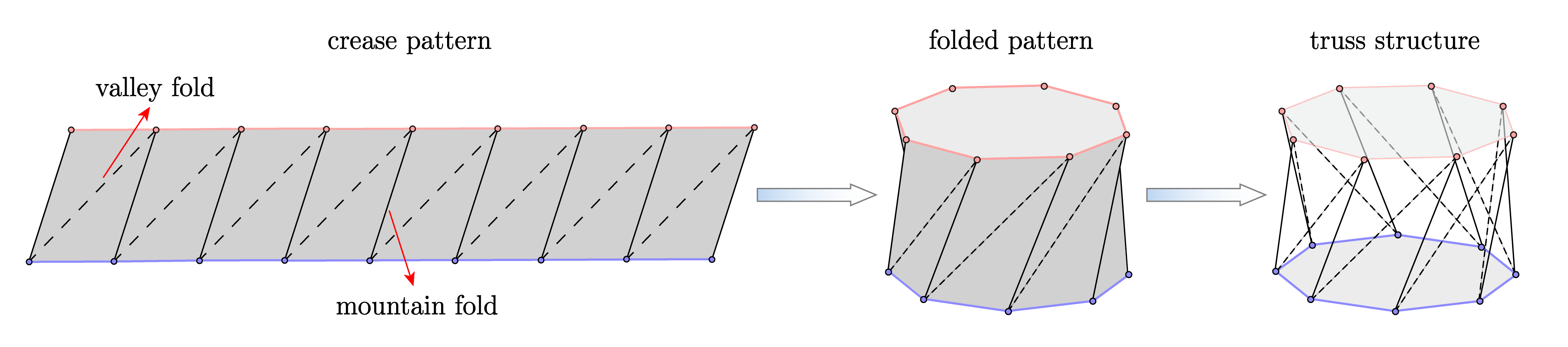}
 \caption{Crease, folded pattern and truss structure of Kresling origami. The solid and dashed lines refer to mountain fold and valley fold, respectively.}
 \end{center}
 \label{fig:crease_pattern}
\end{figure}

The engineering of new structures and metamaterials is increasingly defined by a shift toward exploiting large deformation, nonlinearities, and instabilities to achieve programmable functionality \citep{reis2015perspective,champneys2019happy}. Central to this design challenge are architectures that exhibit axial–twist coupling, where the emergent kinematics of the system link translational displacement to rotational response~\citep{frenzel2017three}. Such coupling provides an interesting mechanism for programming multistability, energy absorption, and tuneable constitutive parameters, such as stiffness and Poisson's ratio~\citep{reid2017geometry,oster2021reentrant,liu2024origami}. One prominent manifestation of this behaviour is found in Kresling origami, a pattern naturally arising from the stress localization in torsional buckling of thin-walled cylindrical shells \citep{kresling2008natural}. The folded configuration and crease pattern of Kresling origami are shown in \autoref{fig:crease_pattern}, where the solid and dashed lines denote mountain and valley folds, respectively. Previous studies modelling the creases as trusses have demonstrated that the stiffness and bistability of Kresling origami can be tuned by tailoring its geometry \citep{PhysRevE.101.063003}. It has also been shown that a more refined geometric description of the Kresling pattern reveals additional interesting mechanics, including a third stable state with significantly increased stiffness. This behaviour arises from crease-level modifications, such as introducing an additional diagonal crease within the triangular panel or replacing the original diagonal crease with an arc~\citep{WANG2023108515,WO2023101941,meloni2021engineering,yang2023volume}.

Although the analysis developed in this work is not limited to Kresling origami, these patterns nevertheless provide a well-established model system for examining how geometry and instability can be used to tailor nonlinear mechanical response. In particular, its functional potential arises from the nonlinear coupling between axial contraction and rotation, which can be harnessed to achieve mechanical responses beyond simple deployment and collapse. This has been demonstrated in coupled Kresling modular structures, where oppositely chiral Kresling units leverage axial–twist coupling to programmable multistability, energy absorption, tuneable stiffness, and self-locking behaviour \citep{LI2020100795}. Beyond the standard cylindrical Kresling pattern, recent studies have explored Kresling derived architectures that extend its nonlinear mechanical response through geometric generalisation, active reconfiguration, and modular assembly. In particular, conical Kresling origami has been developed by relaxing the conventional parallelogram unit-cell geometry to allow free-form quadrilateral cells, thereby enabling tapered/conical folded configurations and an expanded design space \citep{lu2022conical,JIANG2024105796}. Other approaches use magnetic actuation \citep{ZHANG2024105626} or modular chiral assembly \citep{zhao_modular_2025} to achieve tuneable mechanical properties, programmable instability, and reconfigurable shapes.

The preceding examples reflect a broader shift in engineering design, in which nonlinearities are no longer treated as limitations, but are increasingly exploited to achieve programmable functionality \citep{GROH2018394}. For Kresling origami, a detailed understanding of the relationship between geometry and nonlinear mechanics is therefore essential for predicting and tailoring its response. This is particularly challenging because small geometric variations can lead to substantial changes in post-buckling and bifurcation behaviour. Consequently, designing Kresling-based structures requires methods capable of tracking equilibrium paths, switches in stability, and bifurcation points across the relevant design space \citep{doi:10.1098/rspa.2017.0348}. Chain assembly provides a natural route for extending the phenomenology of individual Kresling units to larger structures with targeted mechanical properties. However, this also introduces an additional challenge: the nonlinear equilibrium paths and bifurcation in Kresling chains must be systematically traced to enable inverse design of controllable post-buckling response.

To address these challenges, the present work investigates the bifurcation structure and multistable response of multi-cell Kresling chains, where a generalised framework is developed for $n$-layer assemblies. The geometry and mechanics of the Kresling unit cell are first established in \autoref{sec:KRESLING-BELLOWS-GEOMETRY-MECHANICS}, from which the governing equilibrium equations are derived in \autoref{sec:GOVERNING EQUATIONS OF KRESLING CHAIN}. These equations are then solved over the relevant parameter space to determine the admissible equilibrium branches, associated reaction forces, and energy landscapes for a single unit in \autoref{sec:Single unit and configuration map}. The analysis clarifies the role of the principal geometric parameters in governing stiffness, instability, and energy-absorption capacity, thereby providing a foundation for the rational tailoring of mechanical response. The framework is subsequently extended to multilayer Kresling chains in \autoref{sec:Two-layer chain}-\ref{sec:3-layer}, enabling treatment of the increased dimensionality and interlayer interactions inherent to $n$-layer systems, and yielding a generalised solution for Kresling chains in \autoref{sec:Homogenization solution}.

\section{Kresling kinematics and geometric parameters}
\label{sec:KRESLING-BELLOWS-GEOMETRY-MECHANICS}

\begin{figure}
 \begin{center}
 \includegraphics[width=.95\textwidth]{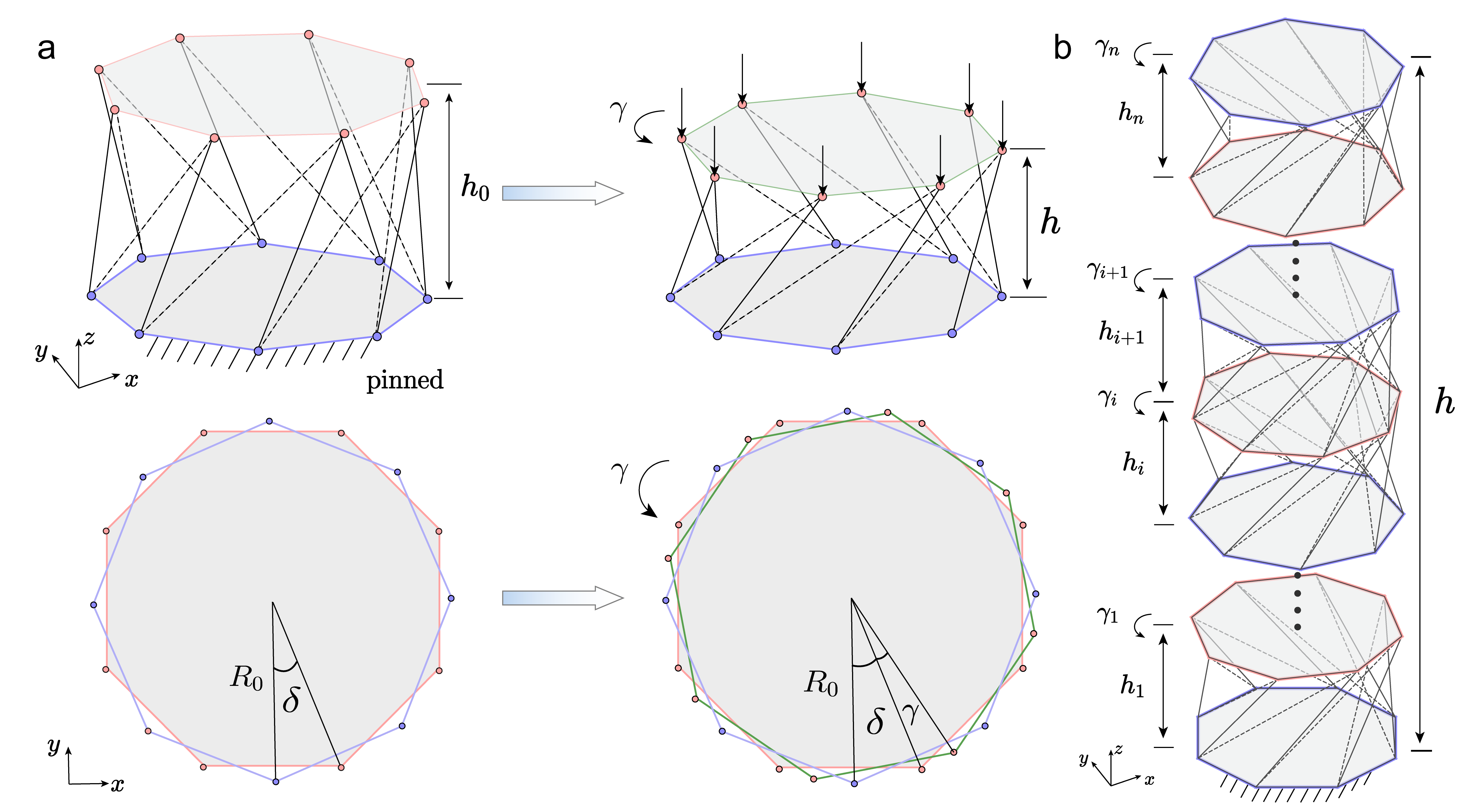} \caption{Definition of geometric parameters and degrees-of-freedom (DOFs) for octagonal Kresling unit cell and chain ($N=8$). (a) Under axial quasi-static displacement control, the Kresling unit cell exhibits a coupled twist behavior. The undeformed geometry is defined by height $h^{(0)}$, base radius ${R^{(0)}}$, and orientation angle $\delta$. The deformed configuration is described by two DOFs: height $h$, and rotational angle $\gamma$. (b) Mechanical model of Kresling chain comprised of $n$ unit cells, where the subscript $i$ denotes the parameters of $i$th individual unit. The total height of the chain, $h$ is employed as the continuation parameter for tracing the equilibrium path. }
 \end{center}
\label{fig:single_coupled}
\end{figure}

Kresling origami is not rigidly foldable; therefore, deploying or collapsing this pattern introduces kinematic incompatibility, which causes stretching and bending of the creases and triangular panels as the structure transitions from its reference configuration to a deformed state~\citep{PhysRevE.101.063003}. Given the complexity in the mechanics of origami, various simplifying assumptions are needed to model how such structures deform. Among these, the bar-hinge model is one of the most widely adopted, simplifying the triangular panels as discrete bars with hinges to capture energy from in-plane deformation and folding along creases \citep{ZANG2024105630,PhysRevLett.110.215501,FILIPOV201726}. The virtual-fold model captures the deformation of bistable Kresling origami by approximating material bending in the triangular panels as rotations about virtual hinge lines, with suitable torsional spring constants assigned to the fold lines \citep{Pagano_2017}. Truss-based modelling has also been widely adopted for Kresling origami, where crease lines or panel edges are represented by axial truss elements. This approach attributes the elastic energy to the elongation and contraction of these elements, and thus does not explicitly account for bending or warping of the triangular panels. Nonetheless, truss-based models have been shown to effectively capture key aspects of Kresling's mechanical response. \citep{FILIPOV201726,10.1115/1.4030158}. Moreover, the truss description of Kresling origami is well-suited for bifurcation analysis, since the energy potential can be expressed as an algebraic system, thereby simplifying the computation of numerical solutions. 

Therefore, this study adopts the truss-based description of Kresling origami to formulate the analysis of its mechanical properties. The relevant geometric parameters and kinematical variables are defined in \autoref{fig:single_coupled}. The vertical and diagonal creases are idealised as deformable truss elements, while the creases of the base panels remain rigid. The Kresling unit (cell), also referred to as a single-layer Kresling, consists of two $N$-sided polygonal base panels with circumferential radius ${R^{(0)}}$, joined by $N$ vertical and diagonal trusses. The initial height, $h^{(0)}$, refers to the undeformed height of the Kresling unit. Similarly, the initial orientation angle, $\delta$, corresponds to the relative angle between the two base panels in the undeformed state. 

\section{Governing equations of a Kresling chain}
\label{sec:GOVERNING EQUATIONS OF KRESLING CHAIN}

Using the truss model, the equilibrium path of a multi-layer Kresling chain can be derived. As shown in \autoref{fig:single_coupled}~(a)-(b), the system is modelled as a chain composed of $n$ units, each with two degrees-of-freedom (DOFs): $h_{i}$ (axial deformation) and $\gamma_{i}$ (rotational deformation), where the subscript $i$ refers to the geometric parameters of the $i$-th unit cell. The total height $h$ of the chain serves as the continuation parameter. The bottom base panel is pinned, and all rotational DOFs are unconstrained. For simplicity, friction and gravity are neglected throughout the analysis.

\begin{figure}
 \begin{center}
  \includegraphics[width=0.95\textwidth]{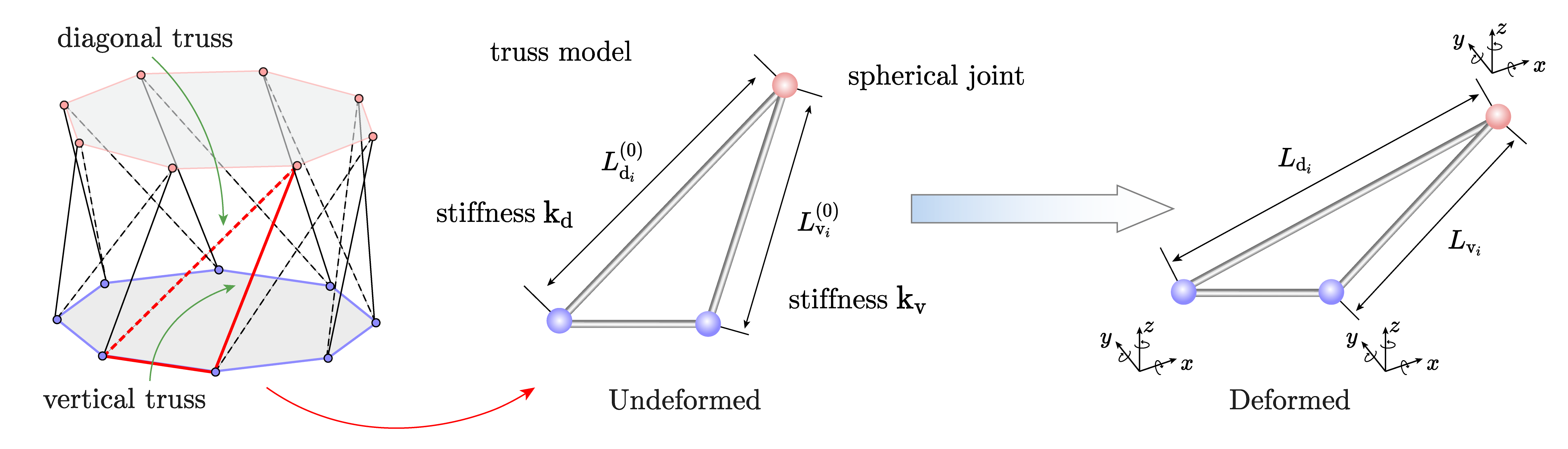}
 \caption{Truss description of Kresling pattern. Deformation mechanism of vertical (v) and diagonal (d) trusses within the triangular facet. Notably, the terms, 'vertical' and 'diagonal', serve only to distinguish the relative positions of trusses, and do not necessarily represent the actual orientations.}
\label{fig:truss_deformation}
 \end{center}
\end{figure}

The kinematics of the truss system, shown in \autoref{fig:truss_deformation}, is first described through the undeformed and deformed lengths of the vertical and diagonal trusses within each unit cell. The undeformed lengths of the vertical ($L^{(0)}_{\mathrm{v}_i}$) and diagonal trusses ($L^{(0)}_{\mathrm{d}_i}$) are expressed in terms of the design geometric parameters as follows:
\begin{subequations}
\begin{align}
    L^{(0)}_{\mathrm{v}_i} =& \sqrt{{h^{(0)}_{i}}^2+2{R^{(0)}}^2-2{R^{(0)}}^2\cos\delta_{i}},\\
    L^{(0)}_{\mathrm{d}_i} =& \sqrt{{h^{(0)}_{i}}^2+2{R^{(0)}}^2-2{R^{(0)}}^2\cos\left(\delta_{i}+\frac{2\pi}{N}\right)}.
\end{align}
\label{eq:UNDEFORMED LENGTH}
\end{subequations}
The deformed lengths of the $i$-th unit cell's vertical ($L_{\mathrm{v}_i}$) and diagonal ($L_{\mathrm{d}_i}$) trusses are derived from the individual height $h_{i}$ and the rotational angle $\gamma_{i}$, which measures the change from the undeformed to the current state. Therefore, we have:
\begin{subequations}
\begin{align}
    L_{\mathrm{v}_i} =& \sqrt{{h_{i}}^2+2{R^{(0)}}^2-2{R^{(0)}}^2\cos(\delta_{i}+\gamma_{i})},\\
    L_{\mathrm{d}_i} =& \sqrt{{h_{i}}^2+2{R^{(0)}}^2-2{R^{(0)}}^2\cos\left(\delta_{i}+\gamma_{i}+\frac{2\pi}{N}\right)}.
\end{align}
\label{eq:DEFORMED LENGTH}
\end{subequations}
Accordingly, the axial deformations $\Delta s_{\mathrm{v}_i}$ and $\Delta s_{\mathrm{d}_i}$, resulting from the elongation and contraction of the vertical (v) and diagonal (d) trusses, can be expressed as follows:
\begin{subequations}
\begin{align}
    \Delta s_{\mathrm{v}_i} &= L_{\mathrm{v}_i}-L^{(0)}_{\mathrm{v}_i}\\
    \Delta s_{\mathrm{d}_i} &= L_{\mathrm{d}_i}-L^{(0)}_{\mathrm{d}_i}.
\end{align}
\label{eq:AXIAL DEFORMATION}
\end{subequations}

Assuming linear elastic behavior, with $\mathrm{k}_{\mathrm{v}}$ and $\mathrm{k}_{\mathrm{d}}$ denoting the axial stiffnesses of the vertical and diagonal trusses, respectively, and noting that all sub-units within each layer experience equal deformation, the internal elastic strain energy of the entire truss system $E_s$, for a $N$-sided polygonal base, can be expressed as:
\begin{equation}
    E_s = \sum_{i=1}^{n} N\left( \frac{1}{2} \mathrm{k}_{\mathrm{v}} {\Delta s_{\mathrm{v}_i}}^2 + \frac{1}{2} \mathrm{k}_{\mathrm{d}} {\Delta s_{\mathrm{d}_i}}^2 \right).
\label{eq:STRAIN_ENERGY}
\end{equation}

We examine the equilibrium paths of a Kresling chain under axial compression. In our analysis, the total height of the chain, $h$, is treated as continuation parameter in the equations. Therefore, the sum of all individual heights must equal the total height:
\begin{equation}
    h-\sum_{i=1}^{n}h_{i} = 0.
\label{eq:CONSTRAINTS}
\end{equation}
The equilibrium path is determined by tracking the local extrema of the total potential energy landscape as the control parameter evolves. An additional height constraint is enforced using the method of Lagrange multipliers, \emph{i.e.} introducing $F$ given by the constraint $E_c$:
\begin{equation}
E_c= F\left(h-\sum_{i=1}^{n}h_{i}\right).
\label{eq:TOTAL POTENTIAL CONSTRAINED 1}
\end{equation}
The Lagrange multiplier $F$ can be interpreted physically as the axial reaction force applied to the cylinder as a function of $h$. 

Moreover, the Kresling chain physically requires all individual heights $h_{i}$ to be positive. To prevent incompatible equilibrium solutions, a linear penalised potential energy is adopted. Specifically, a penalty term $E_p$ should be non-contributory within the admissible region when $h_{i}\geq0$, but impose a large energy barrier when each $h_{i}$ is negative. A simple form of this penalisation term can be written by the following piece-wise function:
\begin{equation}\label{eq:penalty}
E_p = {P}\sum_{i=1}^{n}\langle-h_{i}\rangle_+, \qquad \mathrm{where} \quad \langle-h_{i}\rangle_+=
\left\{
\begin{aligned}
&0, &h_{i}\geq0\\
&-h_{i}, &h_{i}<0,
\end{aligned}
\right .
\end{equation}
where ${P}$, the penalty force, can be interpreted as the contact force between layers, which is set to be high. However, this term exhibits a first-order discontinuity, which creates significant numerical challenges. To address this, instead we use an approximate smooth function to enforce this penalty. This approach preserves the linear penalised energy across the negative domain while ensuring negligible influence in the positive domain. For a given coefficient $\epsilon$, of dimensions of length, in the limit where $\epsilon$ goes to zero, Eq.~\eqref{eq:penalty} can be rewritten as follows: 
\begin{equation}
E_p ={P}\,\lim_{\epsilon\rightarrow0} \sum_{i=1}^{n}  \,\epsilon\,\ln\left(1+e^{-h_{i}/\epsilon}\right).
\label{eq:TOTAL POTENTIAL CONSTRAINED 2}
\end{equation}
Here, as $\epsilon$ decreases, the logarithmic-like energy exhibits an increasing proximity to the piece-wise function. Consequently, we are able to adopt ${P} = 1$ and a sufficiently small $\epsilon$, approximately  $\epsilon\approx 10^{-3}$, to generate the solutions.

By summing the energy terms $E_s$, $E_c$ and $E_p$, the total potential energy can then be written as:
\begin{equation}
E = E_s+ E_c + E_p = \sum_{i=1}^{n} N\left( \frac{1}{2} \mbox{k}_{\mathrm{v}}  {\Delta s_{\mathrm{v}_i}}^2 + \frac{1}{2} \mbox{k}_{\mathrm{d}}  {\Delta s_{\mathrm{d}_i}}^2\right )+F (h-\sum_{i=1}^{N}h_{i}) + {P}\,\lim_{\epsilon\rightarrow0} \sum_{i=1}^{n}  \,\epsilon\,\ln\left(1+e^{-h_{i}/\epsilon}\right).
\label{eq:PENALTY TERM}
\end{equation}
For this system, the equilibrium paths are obtained by applying the principle of stationary potential energy: that is, by taking partial derivatives of the total potential energy.
\begin{subequations}
\begin{align}
&\frac{\partial E}{\partial h_{i}}=N  h_{i}\left(\mbox{k}_{\mathrm{v}}\frac{\Delta s_{\mathrm{v}_i}}{L_{\mathrm{v}_i}}+\mbox{k}_{\mathrm{d}}\frac{\Delta s_{\mathrm{d}_i}}{L_{\mathrm{d}_i}}\right)-{P}\left(\frac{e^{-h_{i}/\epsilon}}{1+e^{-h_{i}/\epsilon}}\right )-F=0,\\
&\frac{\partial E}{\partial \gamma_{i}}=N  {R^{(0)}}^2\left [\mbox{k}_{\mathrm{v}}\frac{\Delta s_{\mathrm{v}_i}}{L_{\mathrm{v}_i}}\sin(\delta_{i}+\gamma_{i})+\mbox{k}_{\mathrm{d}}\frac{\Delta s_{\mathrm{d}_i}}{L_{\mathrm{d}_i}}\sin\left(\delta_{i}+\gamma_{i}+\frac{2\pi}{N}\right)\right ]=0,\\
&\frac{\partial E}{\partial F}=h-\sum_{i=1}^{n}h_{i}=0.
\end{align}
\label{eq:EQUILIBRIUM CRITERION}
\end{subequations}
This algebraic system consists of $(2N+1)$ equations and $(2N+2)$ variables. By solving these equations, we can obtain the individual $h_{i}$, the individual $\gamma_{i}$, the axial reaction force $F$ and the total potential energy $E$ as functions of the continuation parameter $h$.

\section{Bifurcation behaviour and equilibria of Kresling units and their assembly}
\label{sec:RESULTS}
This section investigates the bifurcation structure and multistable behaviour of the Kresling unit and chain under axial compression through numerical solution of the algebraic system defined in Eq.~\eqref{eq:EQUILIBRIUM CRITERION}. Equilibrium branches are tracked using the continuation and bifurcation software AUTO 07P, with total height $h$ taken as the continuation parameter \citep{doedel2007auto}, thereby enabling the systematic identification of bifurcation points and changes in stability. The analysis is restricted to the homogeneous chain, comprising identical unit cells, which, despite its apparent simplicity, exhibits a rich response characterised by multiple bifurcations and coexisting stable equilibria. In certain regimes, the homogeneous chain displays a more complex equilibrium structure than its non-homogeneous counterpart, owing to the symmetry induced by unit-cell uniformity.

Although the total height, $ h $, is used as the continuation parameter in the numerical procedure, the computed equilibrium branches are plotted against the a measure of strain, $\tilde{u}$, defined by 
\begin{equation}
\tilde{u} = -\frac{h-nh^{(0)}}{nh^{(0)}}
\end{equation}
This quantity provides a normalised measure of the axial deformation of the Kresling chain, where the negative sign ensures that larger values of $ \tilde{u} $ correspond to greater axial compression. For the homogeneous Kresling chain considered here, the governing algebraic system indicates that the equilibrium and bifurcation response is dictated primarily by three geometric parameters: the number of polygon sides, $ N $, the initial orientation angle, $ \delta_i $, and the unit-cell height-to-radius ratio, $ h^{(0)}/R^{(0)} $. For most of the analysis that follows, $ N $ is treated as a fixed discrete design choice and the representative aspect ratio $ h^{(0)}/R^{(0)} = 1 $ is adopted as the baseline configuration. This setting enables a clear examination of how $ \delta_i $ shapes the equilibrium structure, bifurcation sequence, and multistable response of the chain.

\subsection{Single unit and configuration map}\label{sec:Single unit and configuration map}

\begin{figure}[pos=ht]
 \begin{center}
 \includegraphics[width=0.95\textwidth]{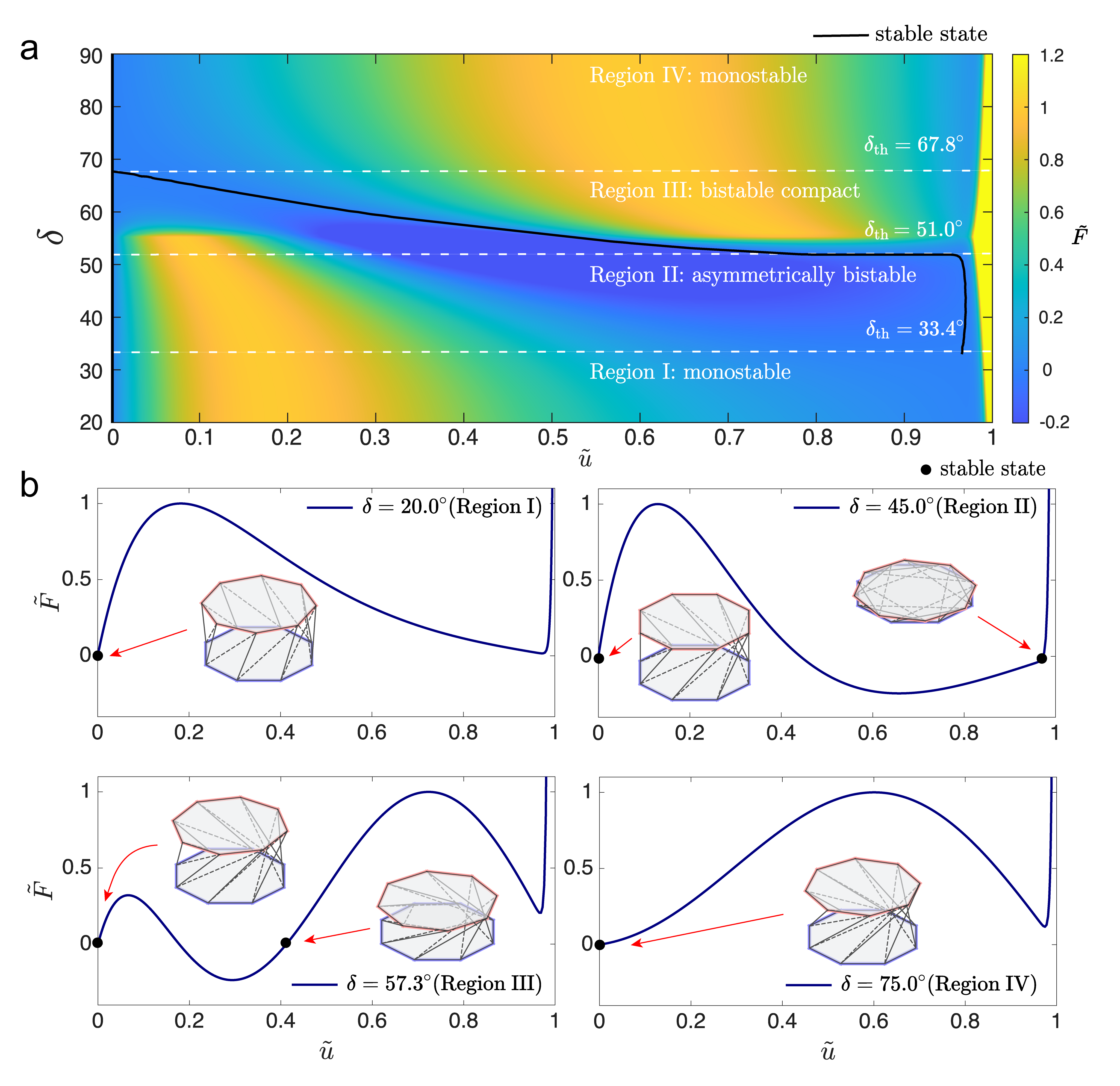}
 \caption{(a) Normalised reaction force curve $\tilde{F}=F/F_{s_{\mathrm{max}}}$ as a function of initial orientation angle $\delta$ , where $F_{s_{\mathrm{max}}}$ refers to the maximum reaction force derived from the truss strain energy $E_s$. The parametric space is partitioned into four distinct regions (Region I–IV) based on $\delta$, each representing different stability characteristic of unit cell. Black lines refer to the stable states, corresponding to local minimum potential energy. (b) Detailed view of normalised reaction force $\tilde{F}$ and deformed configurations of Region I-IV unit cells (from $\delta=20^\circ$ to $\delta=75^\circ$).}
 \label{fig:single_layer}
 \end{center} 
\end{figure}

Since the equilibrium of the Kresling chain is inherited from the constitutive response of an individual unit, we begin by examining the single-unit. \autoref{fig:single_layer}~(a) presents a colour map of the non-dimensional reaction force, $\tilde{F}=F/F_{s_{\mathrm{max}}}$, as a function of $\delta$ and $\tilde{u}$. The results correspond to an octagonal geometry ($N = 8$) with an initial height-to-radius ratio of $h^{(0)}/R^{(0)} = 1$. This single-unit response provides the basis for interpreting the equilibrium structure and multistable behaviour of the full chain. The $\delta$ design space of the unit cell can naturally be divided into four regions (Region I-IV), each defined by its distinct stability characteristics \citep{PhysRevE.101.063003}. In \autoref{fig:single_layer}~(a), the stable equilibrium points for varying values of $\delta$ and $\tilde{u}$ are traced by a black line across the colour map. Meanwhile, \autoref{fig:single_layer}~(b) plots the non-dimensional reaction force $\tilde{F}$ versus $\tilde{u}$ for selected values of $\delta$, effectively presenting slices of \autoref{fig:single_layer}~(a). For Region I, where $\delta <33.4^\circ$, the Kresling unit cell exhibits a single stable state, located at the undeformed state when $\tilde{u}=0$ (\emph{i.e.} $h = nh^{(0)}$).  In Region II where $33.4^\circ<\delta<51.0^\circ$, two asymmetric stable states coexist in the unit cell: the undeformed state ($\tilde{u}=0$) and the nearly collapsed state as $\tilde{u}\rightarrow1$ (\emph{i.e.} $h\rightarrow0_+$). As for Region III, when $51.0^\circ<\delta<67.8^\circ$, bistability is observed, with one stable state at the initial undeformed configuration ($\tilde{u}=0$) and another intermediate state satisfying $0<\tilde{u}<1$ (\emph{i.e.} $0<h/(nh^{(0)})<1$). Notably, this intermediate state is unique as the total potential energy is zero, implying that each truss, whether vertical or diagonal, maintains its undeformed length $L^{(0)}_{\mathrm{v}}$ or $L^{(0)}_{\mathrm{d}}$. For large $\delta>67.8^\circ$ (Region IV), the system exhibits one or two stable states: one at the undeformed configuration ($\tilde{u}=0$) and another possibly at $\tilde{u}<0$ (\emph{i.e.} $h/(nh^{(0)})>1$), which corresponds to axial extension. However, this article focuses on axial compression equilibrium paths; the second (extension) stable state is therefore not considered, and Region IV is defined as monostable.

\begin{figure}[pos=ht]
 \begin{center}
  \includegraphics[width=0.95\textwidth]{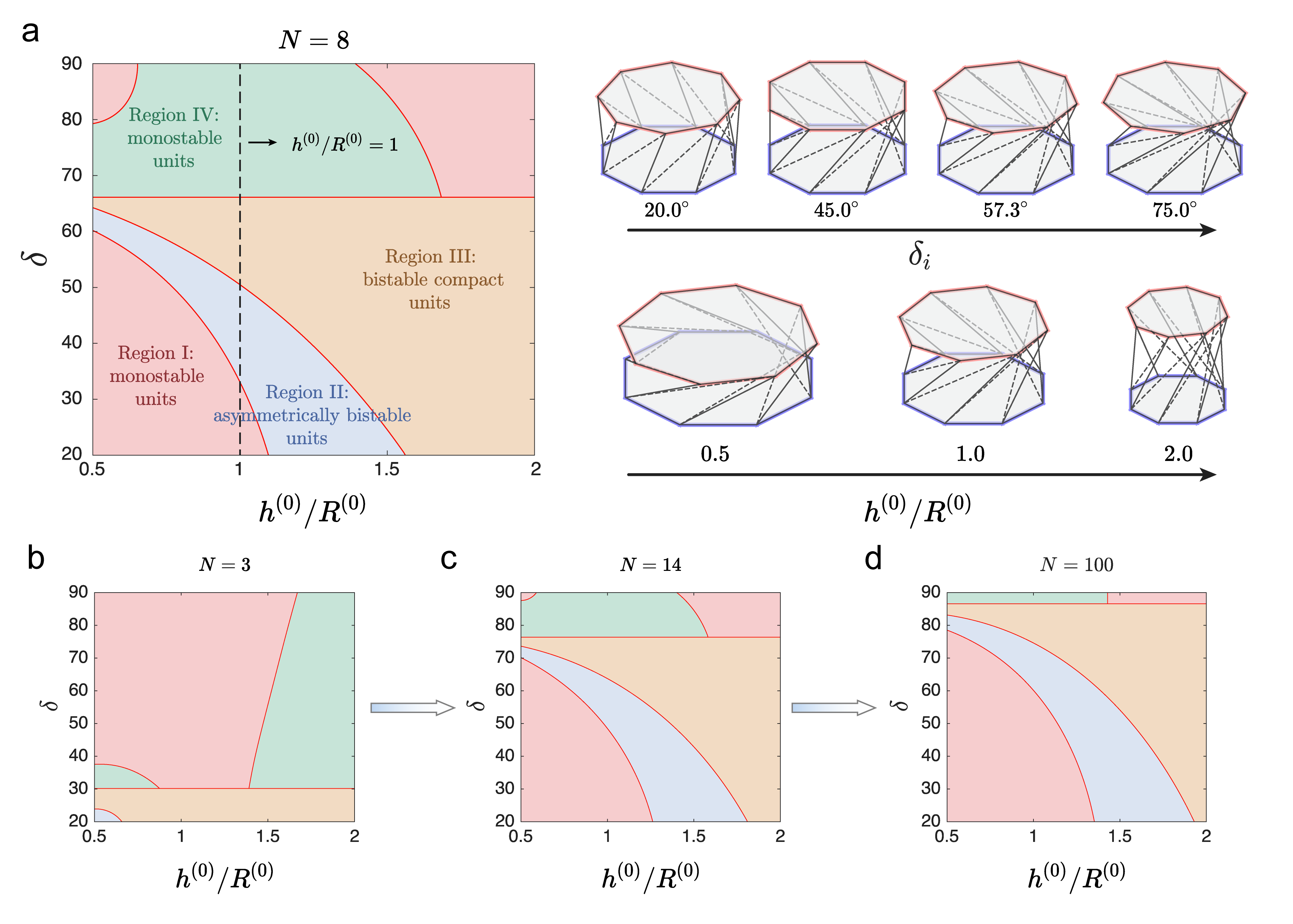}
 \caption{Phase distributions of Region I-IV deforming configurations of Kresling unit cell along two main geometric design space: initial orientation angle $\delta$ and undeformed height-to-radius ratio ${h}^{(0)}/{R^{(0)}}$. (a) Case of octagon base ($N=8$), distribution of Region I-IV and corresponding types of comprised unit cell. The diagrams on the right depict the variations of undeformed configurations along these two geometric design space. (b)-(d) Evolution of Region I-IV distributions from cases of triangle base ($N=3$) to hecatontagon base ($N=100$).}
\label{fig:phase_split}
 \end{center}
\end{figure}

Although the main discussion is conducted for the baseline octagonal geometry with $ h^{(0)}/R^{(0)} = 1 $, it is useful to briefly step beyond this default setting and assess how the remaining geometric design variables modify the single-unit constitutive response. Hence, let us look at the influence of the number of polygon sides, $ N $, and the unit-cell aspect ratio, $ h^{(0)}/R^{(0)} $, since these variations in turn may also shape the equilibrium behaviour of the full chain. To better understand how the distinct bifurcation modes explicitly depend on the unit cell’s geometric parameters, we now present a series of phase diagrams in \autoref{fig:phase_split} that illustrate the distributions of Region I–IV for Kresling unit cells. These regions are defined within the $\delta$ and $h^{(0)}/R^{(0)}$ parameter space. These phase diagrams are constructed from the numerical solutions of the reaction force $\tilde{F}$ profiles (specifically, the position of peak and valley forces), in which the discrete numerical boundaries among phases are smoothed.  We also provide a sequence of plots showing the evolution of these regions from a triangular base ($N=3$) to an hecatontagon base ($N=100$). In \autoref{fig:phase_split}~(a), an inset highlights the physical geometric effect of ${h}^{(0)}/{R^{(0)}}$ on an octagonal chain ($N=8$), representing the cases illustrated in \autoref{fig:single_layer}. The initial orientation angle $\delta_{i}$ ranges from $20^\circ$ to $90^\circ$, and the undeformed height-to-radius ratio ${h}^{(0)}/{R^{(0)}}$ ranges from $0.5$ to $2.0$. 

The phase distribution depends strongly on the number of polygon sides, $N$, especially when $N$ is small. However, as $N$ increases, this dependence diminishes, and the phase transitions stabilise. The evolution of Region I–IV can be divided into two main stages. The first stage occurs at low values of $N$ (3 to 14), where the system is highly sensitive to changes in $N$ and undergoes significant transitions. During this period, as shown in \autoref{fig:phase_split}~(b)–(c), all phase boundaries, including those between Regions I–II, II–III, and III–IV, shift upward continuously within the $\delta_{i}$ design space. Beyond $N\approx14$, the system enters a second stage in which the region distributions begin to exhibits a insensitivity to further increases in $N$, as demonstrated in \autoref{fig:phase_split}~(c)-(d). At this stage, the region's general layout remains consistent. Region I is predominantly located in the bottom-left (low $\delta_{i}$, small ${{h}^{(0)}}/{R^{(0)}}$) and upper-right (high $\delta_{i}$, large ${{h}^{(0)}}/{R^{(0)}}$) portion of the phase diagram. Meanwhile, Region III mainly occupies the central-right part, corresponding to medium $\delta_{i}$ and large ${{h}^{(0)}}/{R^{(0)}}$. Region II bridges between the bottom-left Region I and Region III. Region IV is localised in the upper-central area, representing system with high $\delta_{i}$ and medium ${{h}^{(0)}}/{R^{(0)}}$. 

Among the behaviours observed in these two stages, the three main phase transitions in the parameter space are of particular interest: the Region I-II, II-III, and III-IV/I transitions. These transitions correspond to the three thresholds identified in the single-layer system, as shown in \autoref{fig:single_layer}~(a), and mark changes in the stability or bifurcation characteristics of the chain. When ${{h}^{(0)}}/{R^{(0)}}$ is fixed, all three transition boundaries shift towards larger values of $\delta_i$ as $N$ increases. Similarly, when $\delta_i$ is fixed, the transitions occur at increasingly larger values of ${{h}^{(0)}}/{R^{(0)}}$ as $N$ grows. Notably, all three transitions exhibit a stabilisation trend, where the phase distributions become increasingly similar to further increases in $N$. Overall, these trends provide design insight into how the geometric parameter space can be traversed to tailor the equilibrium paths of these chains.

\subsection{Two-layer chain}
\label{sec:Two-layer chain}

\begin{figure}[pos=ht]
 \begin{center}
  \includegraphics[width=0.95\textwidth]{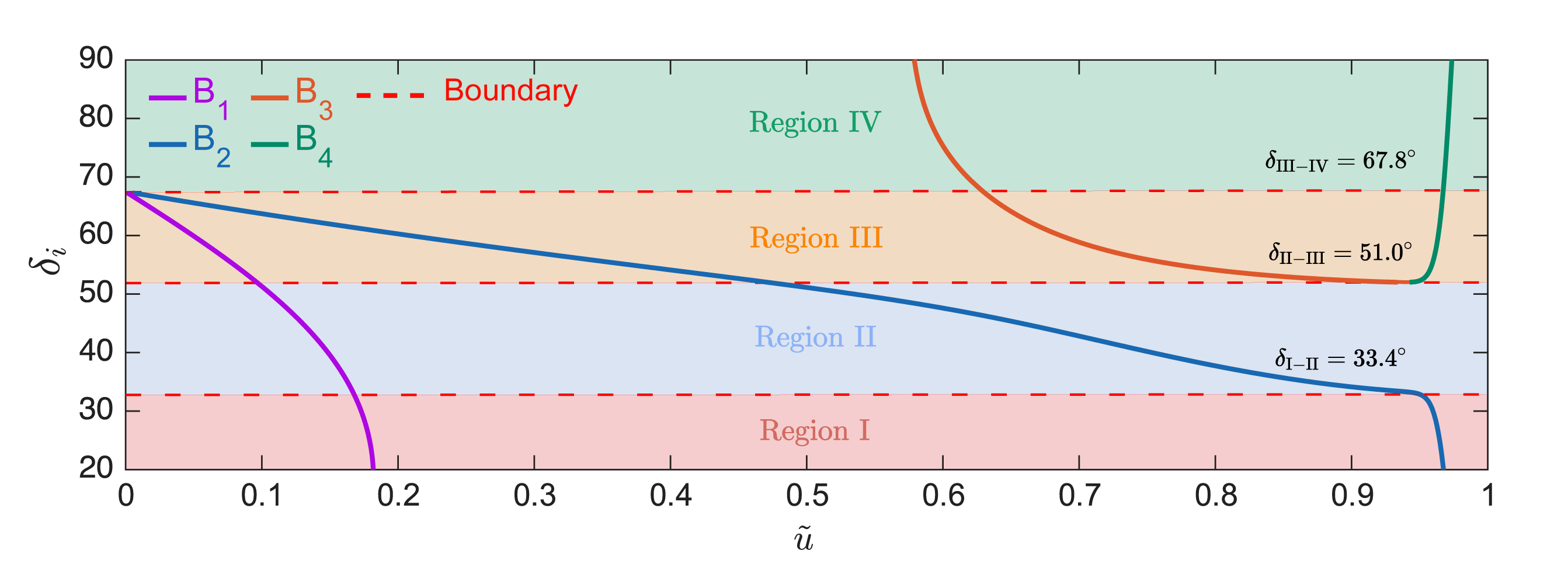}
 \caption{Loci of bifurcation points $\mathrm{B_1}-\mathrm{B_4}$ as the function of initial orientation angle $\delta_{i}$ of comprised unit cells. The parametric space is partitioned into the same four distinct regions (Region I-IV), demonstrating consistency with that of the single unit case.}
\label{fig:split5}
 \end{center}
\end{figure}

\begin{figure}[pos=ht]
 \begin{center}
  \includegraphics[width=0.95\textwidth]{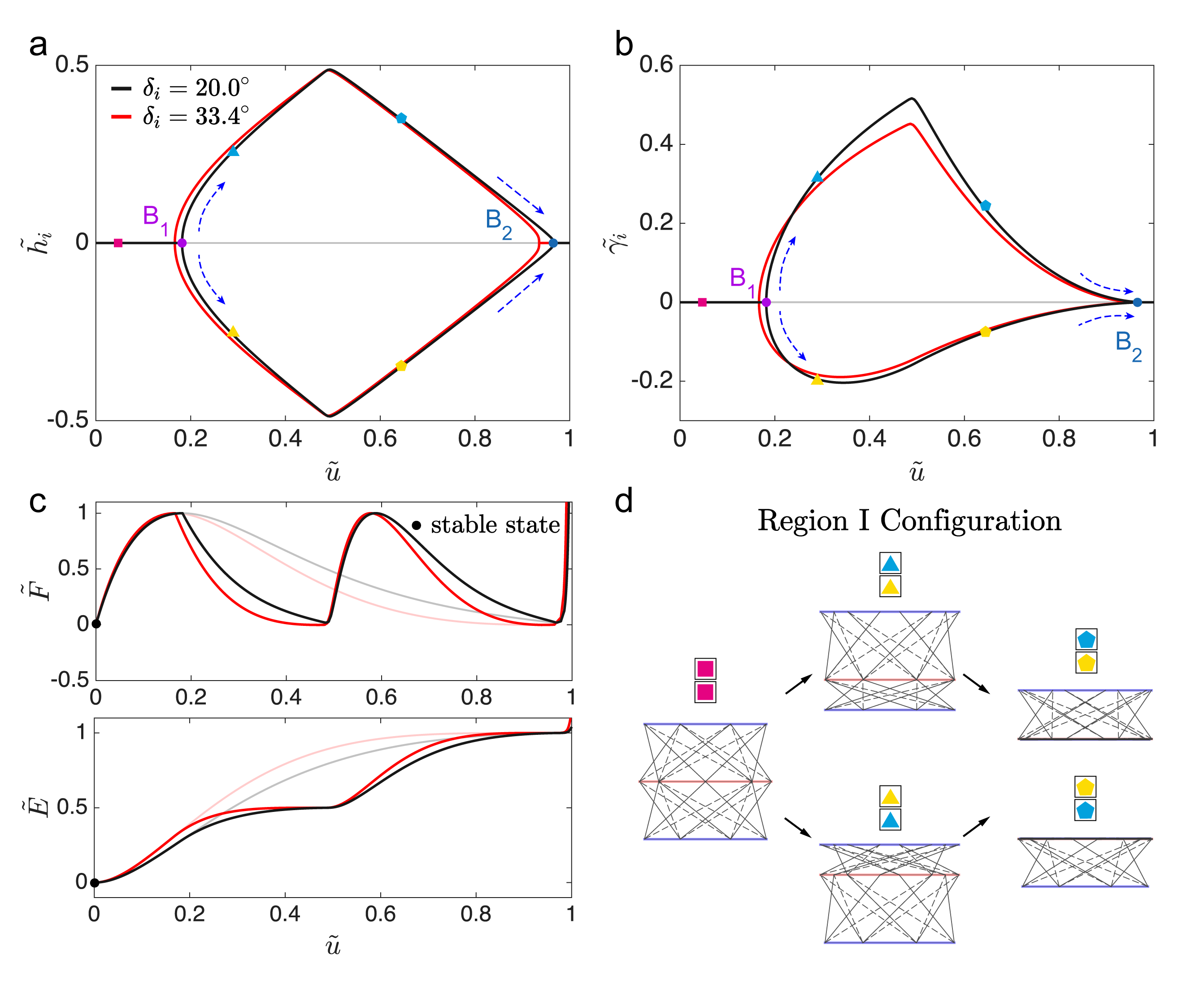}
 \caption{Bifurcation diagrams of the two-layer Kresling chain assembled with identical unit cells. Two cases are compared: Kresling chain comprised of unit cells from Region I ($\delta_{i} = 20.0^\circ$, bifurcation points and stable states are highlighted) and the Region I-II boundary ($\delta_{i} = 33.4^\circ$). Bold and light lines refer to stable and unstable equilibrium path. (a)-(b) Bifurcation diagram described by individual height ${\tilde{h}_{i}}$ and rotational angle ${\tilde{\gamma}_{i}}$ with respect to strain $\tilde{u}$. (c) Normalised reaction force $\tilde{F}=F/F_{s_{\mathrm{max}}}$ and total potential energy $\tilde{E}= E/E_{s_{\mathrm{max}}}$. (d) Schematic of a two-block assembly in a two-layer chain configuration from Region I. The boxed symbols above the chain, such as the blue and yellow triangles, depict the configuration of each unit cell at specific points along the equilibrium path. Each box corresponds to a single layer, with the lowest box representing the bottom layer and the highest box representing the top layer.}
\label{fig:split1}
 \end{center}
\end{figure}

\begin{figure}[pos=ht]
 \begin{center}
  \includegraphics[width=0.95\textwidth]{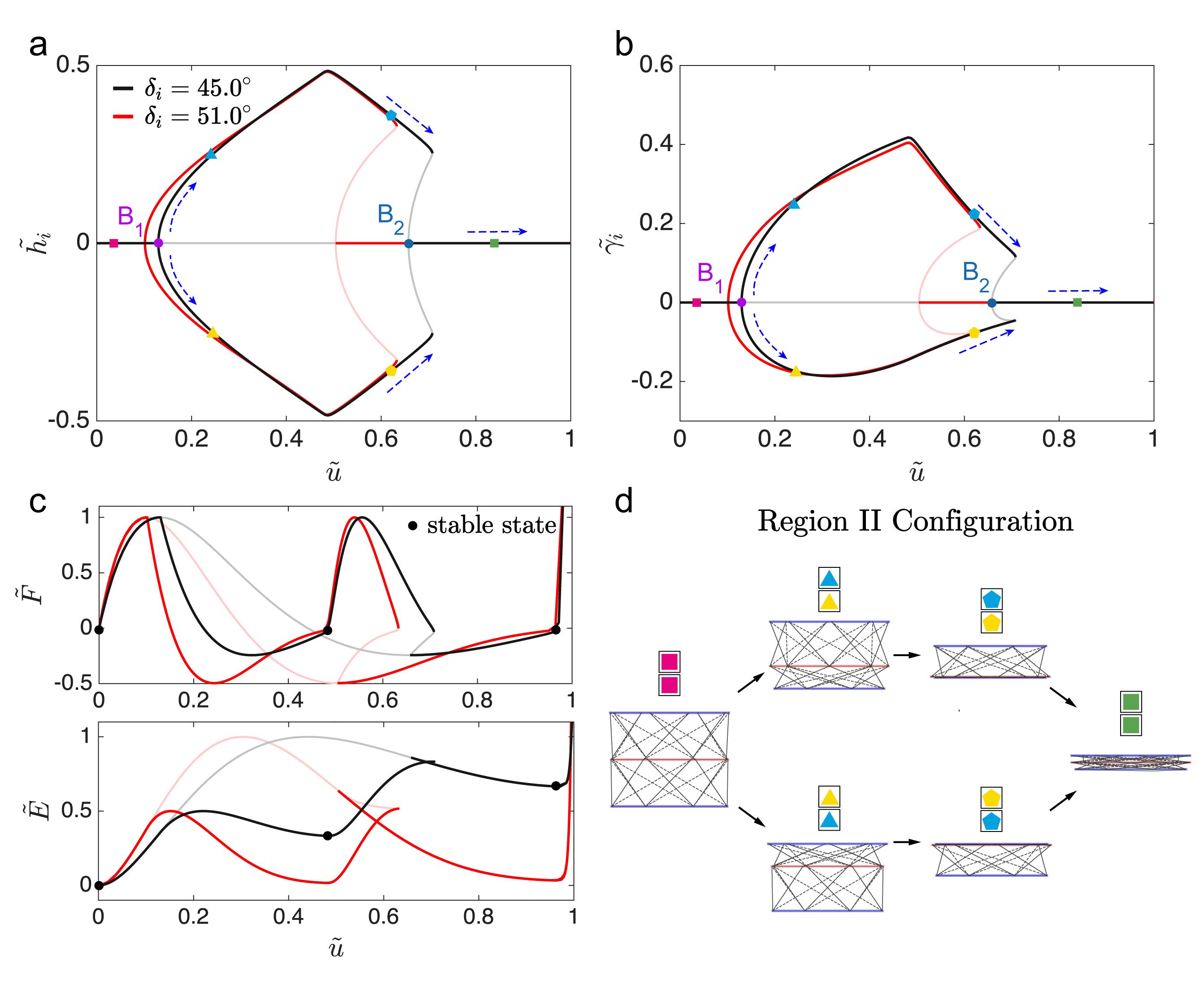}
 \caption{Bifurcation diagrams of the two-layer Kresling chain assembled with identical unit cells. Two cases are compared: Kresling chain comprised of unit cells from Region II ($\delta_{i} = 45.0^\circ$, bifurcation points and stable states are highlighted) and the Region II-III boundary ($\delta_{i} = 51.0^\circ$). Bold and light lines refer to stable and unstable equilibrium. (a)-(b) Bifurcation diagram described by individual height ${\tilde{h}_{i}}$ and rotational angle ${\tilde{\gamma}_{i}}$ with respect to strain $\tilde{u}$. (c) Normalised reaction force $\tilde{F}=F/F_{s_{\mathrm{max}}}$ and total potential energy $\tilde{E}= E/E_{s_{\mathrm{max}}}$. (d) Schematic of a two-block assembly in a two-layer chain configuration from Region II. The boxed symbols above the chain, such as the blue and yellow triangles, depict the configuration of each unit cell at specific points along the equilibrium path. Each box corresponds to one of the layers, with the lower box representing the bottom layer and the higher box representing the top layer.}
\label{fig:split2}
 \end{center}
\end{figure} 

\begin{figure}[pos=ht]
 \begin{center}
  \includegraphics[width=0.95\textwidth]{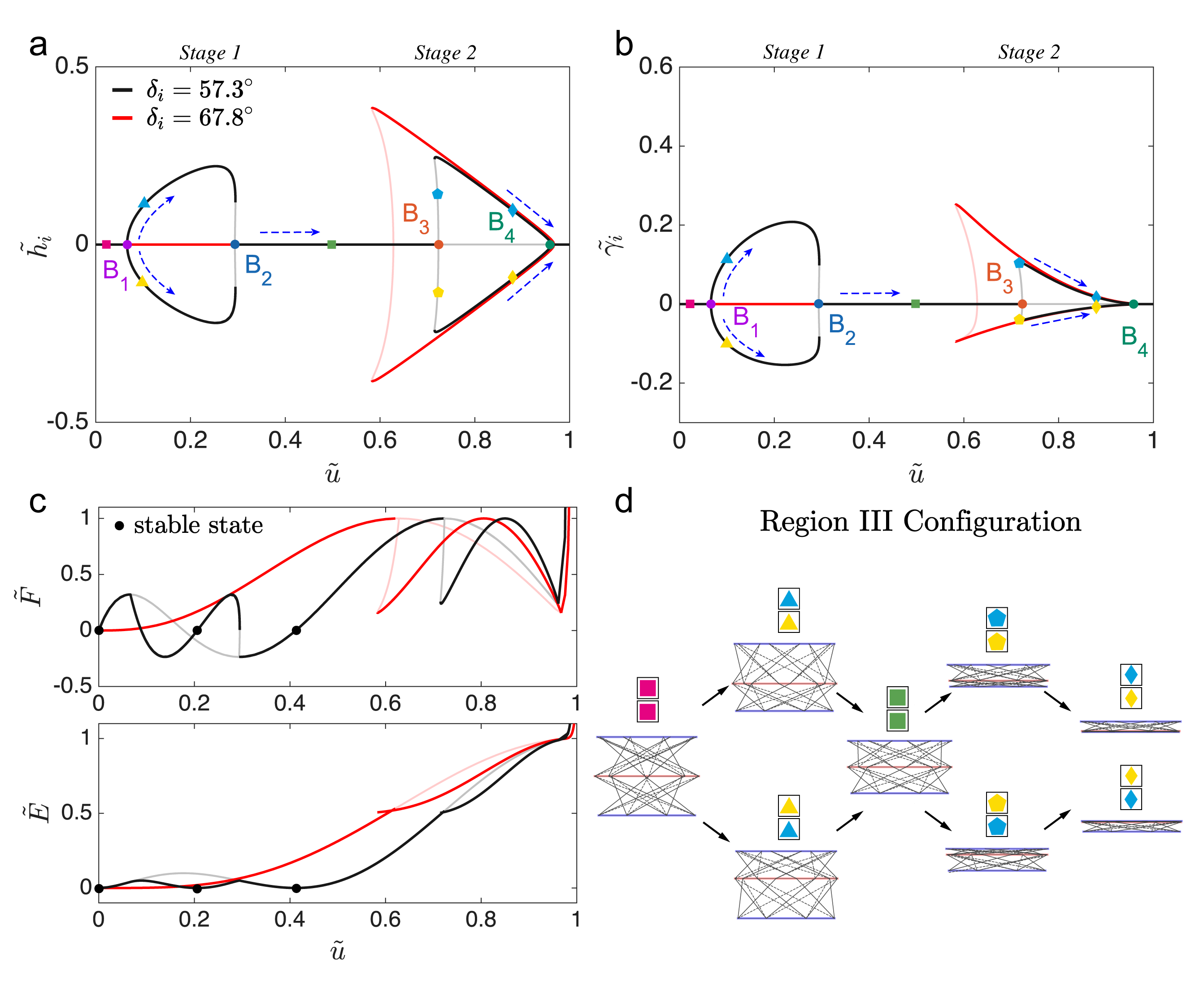}
 \caption{Bifurcation diagrams of the two-layer Kresling chain assembled with identical unit cells. Two cases are compared: Kresling chain comprised of unit cells from Region III ($\delta_{i} = 57.3^\circ$, bifurcation points and stable states are highlighted) and the Region III-IV boundary ($\delta_{i} = 67.8^\circ$). Bold and light lines refer to stable and unstable equilibrium. (a)-(b) Bifurcation diagram described by individual height ${\tilde{h}_{i}}$ and rotational angle ${\tilde{\gamma}_{i}}$ with respect to strain $\tilde{u}$. (c) Normalised reaction force $\tilde{F}=F/F_{s_{\mathrm{max}}}$ and total potential energy $\tilde{E}= E/E_{s_{\mathrm{max}}}$. (d) Schematic of a two-block assembly in a two-layer chain configuration from Region III. The boxed symbols above the chain, such as the blue and yellow triangles, depict the configuration of each unit cell at specific points along the equilibrium path. Each box corresponds to one of the layers, with the lower box representing the bottom layer and the higher box representing the top layer.}
\label{fig:split3}
 \end{center}
\end{figure}

\begin{figure}[pos=ht]
 \begin{center}
  \includegraphics[width=0.95\textwidth]{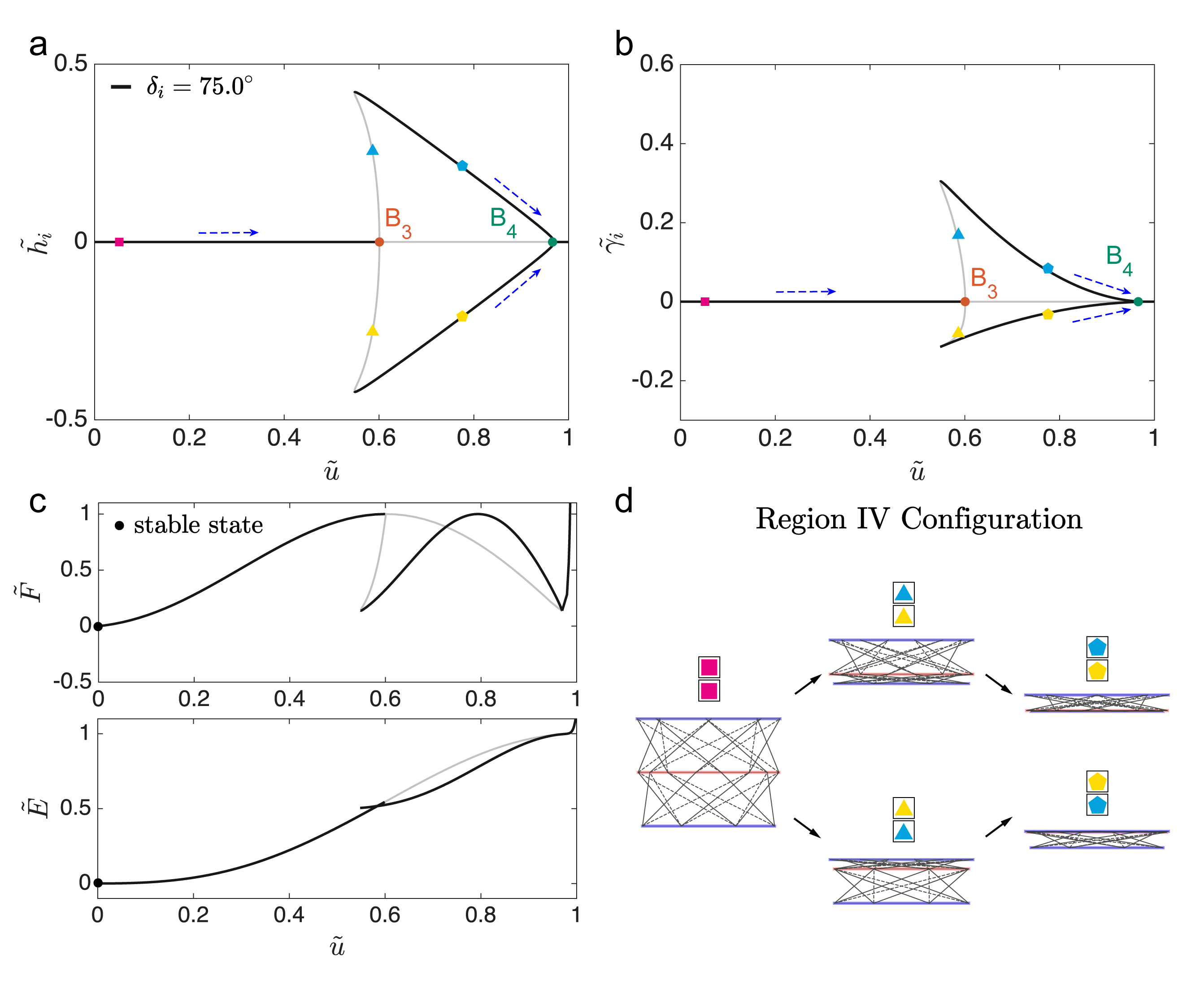}
 \caption{Bifurcation diagrams of the two-layer Kresling chain assembled with identical unit cells from Region IV ($\delta_{i} = 75.0^\circ$, bifurcation points and stable states are highlighted). Bold and light lines refer to stable and unstable equilibrium. (a)-(b) Bifurcation diagram described by individual height ${\tilde{h}_{i}}$ and rotational angle ${\tilde{\gamma}_{i}}$ with respect to strain $\tilde{u}$. (c) Normalised reaction force $\tilde{F}=F/F_{s_{\mathrm{max}}}$ and total potential energy $\tilde{E}= E/E_{s_{\mathrm{max}}}$. (d) Schematic of a two-block assembly in a two-layer chain configuration from Region IV. The boxed symbols above the chain, such as the blue and yellow triangles, depict the configuration of each unit cell at specific points along the equilibrium path. Each box corresponds to one of the layers, with the lower box representing the bottom layer and the higher box representing the top layer.}
\label{fig:split4}
 \end{center}
\end{figure}

Following the characterisation of the single-unit response, we now consider the simplest assembly, a chain composed of two identical Kresling units. The same baseline geometry, defined by $ N =8 $ and $ h^{(0)}/R^{(0)} = 1 $, while varying $ \delta_i $, is retained. In contrast to the isolated unit, the two-layer chain can exhibit symmetry-breaking bifurcations because the additional internal DOF allows the two units to deform unequally. This motivates expressing each unit state relative to the chain-average configuration, thereby filtering out the mean compression and isolating the nonuniform mode associated with symmetry breaking. To quantify departures from homogeneous deformation within the chain, we introduce layer-wise measures of the deviations in unit height and rotational angle. Let $h_{\mathrm{avg}}=\sum_{j=1}^{n} h_j/n$ denote the average height of the unit cells. The normalised height deviation of the $i$th unit is then defined as
\begin{equation}
\tilde{h}_i=\frac{h_i-h_{\mathrm{avg}}}{R^{(0)}}.
\end{equation}
Hence, $ \tilde{h}_i = 0 $ represents a homogeneously deformed chain, whereas nonzero values quantify layer-to-layer non-uniformity. Moreover, along the minimum-energy path, the rotational angle is in bijective correspondence with the unit height, so that $ \gamma_i $ can be expressed as a function of $ h_i $. The corresponding angular deviation is therefore defined as
\begin{equation}
\tilde{\gamma}_i=\gamma_i\left({h_i}\right)-\gamma_i\left({h_{\mathrm{avg}}}\right).
\end{equation}
This quantity measures the difference between the rotation of the $ i $th unit and the rotation associated with the average unit height. Hence, $ \tilde{\gamma}_i $ is not an independent variable, but rather the rotational counterpart of the local height deviation. Together, $ \tilde{h}_i $ and $ \tilde{\gamma}_i $ provide meaningful measures of of departures from a spatially uniform deformation state, and will be used below to identify symmetry breaking and localised deformation within the chain.

We now focus on how $\delta_i$ affects the bifurcation behaviour of the chain's equilibrium path. To capture this dependence, \autoref{fig:split5} shows the loci traced by the bifurcation and coalescence points $\mathrm{B_1}$, $\mathrm{B_2}$, $\mathrm{B_3}$, and $\mathrm{B_4}$ with changing $\delta_i$, illustrating their evolution across the design space. \autoref{fig:split5} also overlays the region classification for a single unit, which is important because the constitutive behavior of the individual units strongly influences the overall chain response. The critical values of $\delta_i$ defining these region boundaries, introduced earlier, are $\delta_{\mathrm{I-II}}=33.4^\circ$, $\delta_{\mathrm{II-III}}=51.0^\circ$, and $\delta_{\mathrm{III-IV}}=67.8^\circ$.

~\autoref{fig:split1} to \autoref{fig:split4} present a series of bifurcation diagrams for $\delta_{i}$ values ranging from $20^\circ$ to $75^\circ$.  Each figure illustrates the chain’s response when built from unit cells representing one of the four distinct regions. Additionally, each figure includes a case exactly at each transitional $\delta_{i}$ value, highlighting how the chain’s bifurcation behaviour changes at the boundary between two regions defined by the stability criteria of the unit cell. Overall, stable equilibrium paths are shown with bold lines, while unstable paths are depicted with light lines. As derived in \autoref{sec: appendix B}, stability is determined by the positive definiteness of the projected Hessian matrix $H_p$ of the potential energy: namely $H_p \succ 0$ for stable equilibrium and $H_p \preceq 0$ for unstable equilibrium. The positive definiteness of $H_p$ ($H_p \succ 0$) signifies local convexity of the potential energy function, ensuring stable progression along the equilibrium path. Conversely, for unstable equilibrium ($H_p \preceq 0$), characterised by concavity, the system will experience snap-through to the nearest stable branch with lower potential energy, where the direction of snap-through is governed by how load is applied (e.g., force-control, displacement-control). In this analysis, we assume the equilibrium paths follow a quasi-static process, neglecting nonlinear dynamical effects. 

\autoref{fig:split1} displays the bifurcation response of a chain constructed from Region I unit cells, along with the I--to--II transition. Notice that the bifurcation and coalescence points are marked as $\mathrm{B}_{1}$ and $\mathrm{B}_{2}$. As shown in \autoref{fig:split1}~(a)-(b), the bifurcation diagram for $\delta_i=20.0^\circ$, representative of Region I, is initially characterised by synchronous and uniform deformation of the two unit cells. This behaviour continues up to the super-critical pitchfork bifurcation point $\mathrm{B_1}$, where the primary path bifurcates into two stable secondary branches. Here, the primary equilibrium path refers to the configuration in which all unit cells share the same deformation. Along the secondary branches, one unit cell gradually collapses while the other undergoes release. Here, release refers to the process in which the unit cell expands. After the first unit cell has fully collapsed, the second unit cell continues along the stable branch path until both unit cells return to the primary path at $\mathrm{B_2}$, which we refer to as a coalescence point. The transition from Region I to Region II occurs when $\delta_i$ exceeds $33.4^\circ$. At this stage, the position of $\mathrm{B}_2$ point gradually shifts toward lower $\tilde{u}$ and the transitions from stable to unstable coalescence. $\tilde{F}$ and $\tilde{E}$ plots (\autoref{fig:split1}~(c)) indicate that the undeformed configuration is the only stable state for Region I case. In \autoref{fig:split1}~(d), the sequential configurations of corresponding unit cell clearly reflects this whole bifurcation process. 

The response of chains constructed from Region II unit cells follows the same initial behaviour as chains constructed from Region I units. As shown in \autoref{fig:split2}~(a)-(b), both unit cells initially undergo uniform deformation up to the supercritical pitchfork bifurcation point $\mathrm{B_1}$, after which one of the unit cells begins to progressively collapse. However, if displacement-controlled, unlike the chain constructed from Region I units, the bifurcated branches during this collapse process lose stability, causing a sudden snap-through towards the primary equilibrium path and a corresponding precipitous drop in load-bearing capacity. The transition from Region II to Region III happens as the $\delta_i$ exceeds $51.0^\circ$. During this stage, as indicated by \autoref{fig:split5}, two additional bifurcation points $\mathrm{B_3}$ and $\mathrm{B_4}$ appear, marking the emergence of as second bifurcation stage. And the loci of $\mathrm{B_1}$ and $\mathrm{B_2}$ in Region II continue to shift toward lower $\tilde{u}$. The snap-through can also be reflected in $\tilde{F}$ and $\tilde{E}$ plots in \autoref{fig:split2}~(c), where discontinuities indicate nonlinear dynamic responses and potential energy dissipation. In addition to the undeformed stable state, two further stable states are discernible for Region II case: one corresponding to the full collapse of a single unit cell, and the other to the compact state in which both unit cells have collapsed. The corresponding bifurcated configurations are demonstrated in \autoref{fig:split2}~(d).

Unlike chains constructed from Region I or Region II unit cells, the chain constructed from Region III unit cells, shown in \autoref{fig:split3}~(a)-(b), exhibits two distinct stages of bifurcation behavior, denoted as Stage 1 and 2. This includes two bifurcation points ($\mathrm{B_1}$ and $\mathrm{B_3}$) and two coalescence points ($\mathrm{B_2}$ and $\mathrm{B_4}$), where the branches merge. The first bifurcation point, $\mathrm{B_1}$, is a supercritical pitchfork bifurcation that marks the onset of asymmetric folding. Beyond $\mathrm{B_1}$, the equilibrium path splits into two stable branches, which soon merge back into the primary path at $\mathrm{B_2}$. The second bifurcation point, $\mathrm{B_3}$, is a subcritical pitchfork bifurcation, leading to two unstable branches and inducing a snap-through instability along the equilibrium path, which triggers the collapse of a single unit cell. The boundary between Region III and Region IV is located at $\delta_i=67.8^\circ$. At this point, the branches connecting $\mathrm{B_1}$ and $\mathrm{B_2}$ collapse onto the primary path, causing these bifurcations to disappear. The $\mathrm{B_3}$ and $\mathrm{B_4}$ points persist, and as $\delta_{i}$ increases, they gradually move further apart along $\tilde{u}$, as illustrated in \autoref{fig:split5}. As seen in the $\tilde{F}$ and $\tilde{E}$ plots indicated by \autoref{fig:split3}~(c), the deformation process for Region III exhibits three stable states, including the initial undeformed state. The second stable state is found on the branches emerging from the first bifurcation point ($\mathrm{B_1}$) while the third stable state follows the primary path after the reconnection of the first bifurcated branches and has a higher energy barrier for compression compared to extension. Corresponding bifurcated configurations are illustrated in \autoref{fig:split3}~(d).

The chain constructed from Region IV unit cells, shown in \autoref{fig:split4}~(a)-(b), exhibits similar behaviour as the second stage of chains constructed from Region III unit cells. After the uniform deformation stage, the equilibrium path arrives at the subcritical bifurcation point $\mathrm{B_3}$, where two unstable branches emerge. The label $\mathrm{B_3}$ is retained here, as it originates from the evolution of the $\mathrm{B_3}$ point observed in Region III (see \autoref{fig:split5}). At this point, if displacement-controlled, one unit cell snaps into a compact state, while the other undergoes a sudden release. Afterwards, the previously released unit cell experiences stable compression until the two branches merge back into the primary path. Consistent with the second stage of chains constructed from Region III unit cells, no stable states exist other than the initial undeformed state. And the whole bifurcation process can be illustrated in \autoref{fig:split4}~(d).

This comparison between isolated unit cell and that of two-layer chains constructed from each region demonstrates the significant impact of series assembly on global stability and bifurcation patterns. The four regions are defined solely by the constitutive properties of the individual unit cell: Regions I and IV correspond to monostable unit cells, while Regions II and III correspond to bistable unit cells. When two unit cells are assembled into a chain, the additional unit introduces extra DOFs, allowing non-uniform configurations that are impossible in the single-unit case. As a result, chains constructed from Region I or Region IV unit cells remain monostable at the chain level, whereas chains built from Region II or Region III unit cells acquire an additional stable state, resulting in three stable states overall. This enriched stability landscape is reflected in the chain’s bifurcation: chains from Region I unit cells (monostable) display a supercritical pitchfork bifurcation; chains from Region II unit cells (asymmetrically bistable) exhibit unstable re-merging during the first bifurcation stage; chains from Region III unit cells (bistable compact) show two bifurcation stages involving both supercritical and subcritical branching; and chains from Region IV unit cells (monostable with possible extension-stable state) exhibit a single subcritical branching point. These observations demonstrate that while single-unit classification guides the selection of unit-cell type, assembling them into a two-layer chain modifies the overall mechanical response by introducing additional DOFs and, in Regions II and III, extra stable states.

\subsection{Three-layer chain}
\label{sec:3-layer}

\begin{figure}[pos=p]
 \begin{center}
  \includegraphics[width=0.95\textwidth]{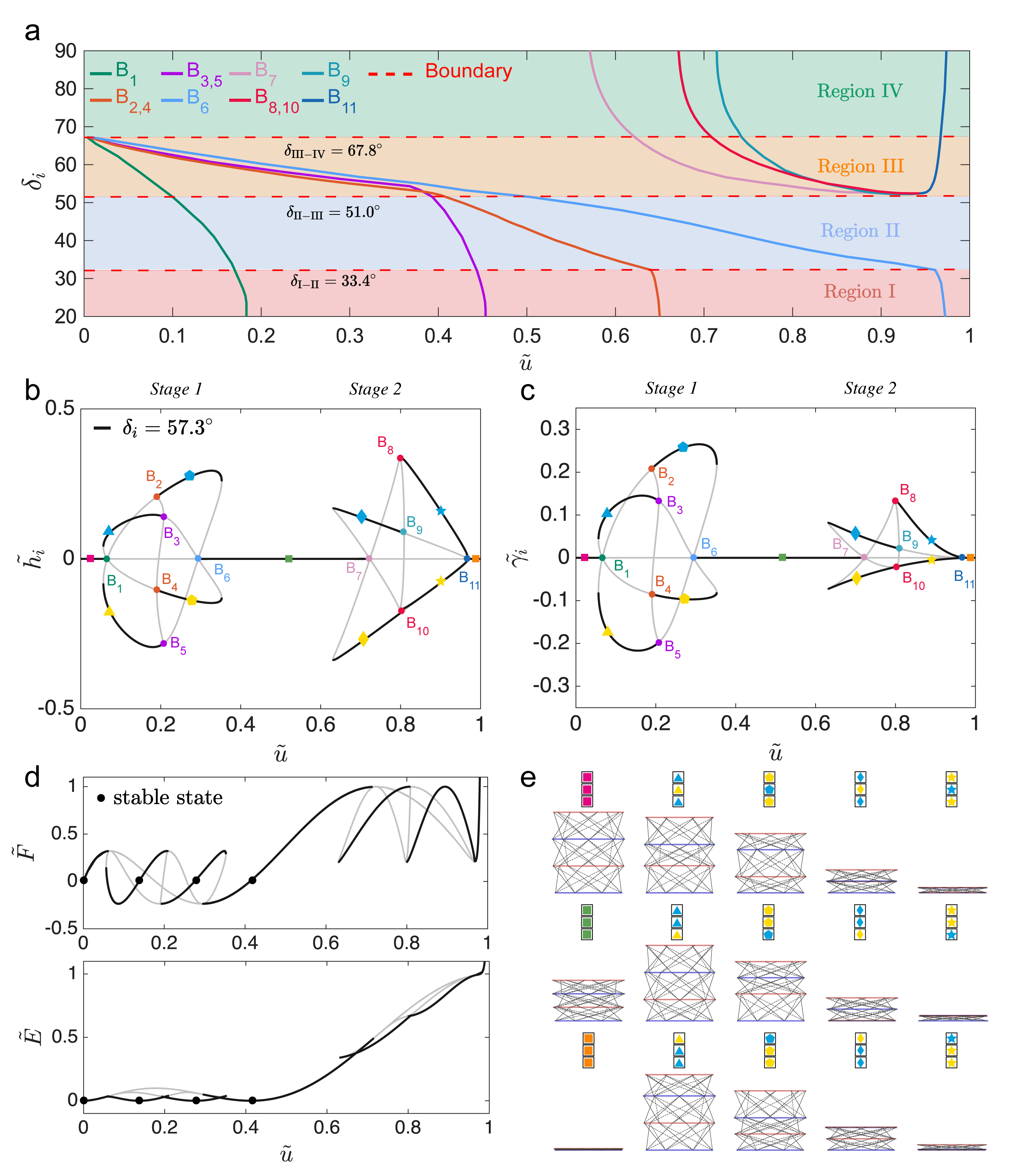}
 \caption{Bifurcation diagrams of the three-layer Kresling chain assembled with identical unit cells from Region III ($\delta_{i} = 57.3^\circ$, bifurcation points and stable states are highlighted). Bold and light lines refer to stable and unstable equilibrium. (a) Loci of bifurcation points $\mathrm{B_1}-\mathrm{B_{11}}$ as the function of unit cell initial orientation angle $\delta_{i}$. The parametric space is partitioned into the same four distinct regions (Region I-IV). (b)-(c) Bifurcation diagram described by individual height ${\tilde{h}_{i}}$ and rotational angle ${\tilde{\gamma}_{i}}$ with respect to strain $\tilde{u}$. (d) Normalized reaction force $\tilde{F}=F/F_{s_{\mathrm{max}}}$ and total potential energy $\tilde{E}= E/E_{s_{\mathrm{max}}}$. (e) Schematic of a three-block assembly in a three-layer chain configuration from Region III. The boxed symbols above the chain, such as the blue and yellow triangles, depict the configuration of each unit cell at specific points along the equilibrium path. Each box corresponds to one layer, with the lowest box representing the first layer and the highest box representing the third layer. }
\label{fig:3-layer}
 \end{center}
\end{figure}
 
To understand how additional DOFs affect the equilibrium and bifurcation paths, we next examine a three-layer chain constructed from Region III unit cells, extending the preceding two-layer analysis toward the general $n$-layer Kresling chain. We select Region III because chains built from these unit cells display the full set of bifurcation points (\autoref{fig:3-layer}~(a)), offering the most comprehensive insight into the system's bifurcation. This analysis provides a foundation for understanding and generalising the behaviour of Kresling chains with an arbitrary number of layers, which will be addressed in the following section.  

As for three-layer system, one notable aspect is the location and number of the bifurcation points ($\mathrm{B_1}$-$\mathrm{B_{11}}$), as demonstrated in \autoref{fig:3-layer}~(a). The positions of the bifurcation points along the primary equilibrium path, $\mathrm{B_1}$, $\mathrm{B_6}$, $\mathrm{B_7}$, and $\mathrm{B_{11}}$ in the three-layer system, coincide with $\mathrm{B_1}$, $\mathrm{B_2}$, $\mathrm{B_3}$, and $\mathrm{B_4}$ in the two-layer chain. In both cases, these points mark the start and end of each bifurcation stage, representing the bifurcation and coalescence of the primary branch. The main differences arise along the secondary and tertiary equilibrium paths: the three-layer chain exhibits a greater number and density of bifurcation points, determined by the possible ways to partition the unit cells into two or three distinct states. Specifically, two possible partitions of secondary paths emerge: one of the units adopts lower state while the other two occupy higher state, and vice versa. And for tertiary paths, only one partition exists: the three units occupy three distinct states. Consequently, these partitions lead to four secondary branch points ($\mathrm{B_2}-\mathrm{B_5}$) of Stage 1 and three ($\mathrm{B_8}-\mathrm{B_{10}}$) of Stage 2. 

The corresponding bifurcation diagram for the three-layer case constructed from Region III unit cells is demonstrated in \autoref{fig:3-layer}~(b)-(c). At each stage, the equilibrium path splits into a primary branch and two pairs of secondary branches, which later remerge with the primary path at higher load levels. The secondary branches originate from the initial bifurcation and represent the transition where unit cells lose uniform deformation and adopt the partitions of two distinct states mentioned above. Beyond this, similarly, tertiary branches emerge from further bifurcation of the secondary branches, resulting in three distinct deformation states. 

The differences in bifurcation between the two-layer and three-layer systems arise directly from the structure of the governing equations. Because the system is algebraically symmetric, these equations remain unchanged when the coupled DOFs across different layers (\emph{i.e.} individual $h_{i}$ and $\gamma_{i}$) are interchanged. This underlying symmetry allows the two-layer system to be recovered as a constrained reduction of the three-layer formulation along the secondary branches, in which selected $h_{i}$ and $\gamma_{i}$ are tied together, \emph{i.e.}
\begin{equation}
\left \{
\begin{aligned}
H_{\mathrm{constraint}} &= h_{3} - h_{j}=0 \\
\Gamma_{\mathrm{constraint}} &=\gamma_{3} - \gamma_{j}=0 
\end{aligned}
\right .
\quad \text{for  } j = 1 \text{  or  } 2.
\end{equation}
This is to say, under the constraint of identical coupled DOFs between the third unit and one of the other two, the three-layer system can then be simplified to another equivalent two-layer system. 

Guided by this analysis, we now revisit the bifurcation diagram of the three-layer chain shown in \autoref{fig:3-layer}. During the initial loading phase, all three unit cells deform identically along the primary equilibrium path. At the first bifurcation point ($\mathrm{B_1}$), the system branches into secondary equilibrium paths, indicating the onset of localised deformation. Here, two unit cells retain similar behaviour while the third diverges, resulting in two distinct states among the three units. Further along, tertiary equilibrium branches, though unstable, mark the point where the remaining two units also break symmetry, leading to a fully non-symmetrised state in which each unit cell exhibits a unique configuration. Due to the instability of the tertiary branches, if displacement-controlled, the system undergoes snap-through transitions, jumping between secondary paths. This symmetry-breaking and snap-through occur at both main bifurcation stages, from $\mathrm{B_1}$ to $\mathrm{B_6}$ and from $\mathrm{B_7}$ to $\mathrm{B_{11}}$. As illustrated in \autoref{fig:3-layer}~(d), the number of stable states increases: there are four stable states in the first bifurcation stage, one more than in the two-layer chain constructed with the same Region III units. Detailed configurations of corresponding unit cells along equilibrium paths are displayed in \autoref{fig:3-layer}~(e).

\section{Generalised solution for $n$-layer Kresling chain}
\label{sec:Homogenization solution}

While the governing algebraic equations (Eq.\eqref{eq:EQUILIBRIUM CRITERION}) theoretically determine the equilibrium paths and bifurcations of multi-layer Kresling chains, practical numerical solutions are complicated by the system's inherent symmetry. This symmetry leads to a singular Jacobian when approaching branching and coalescence points, which in turn causes standard numerical continuation methods to fail. To address this challenge, we introduce a generalisation approach within this section.

\subsection{Kresling with multiple units}
\label{sec:Kresling with multiple units}

Observations from the three-layer system show that secondary bifurcated paths arise when, among the three unit cells, the two that share identical deformation following the bifurcation of one of the units themselves undergo a bifurcation, leading to further symmetry breaking. This insight, as detailed in \autoref{sec:3-layer}, forms the basis for our generalised method. We now extend this approach to the $n$-layer assembled structure, analysing how this symmetry-breaking mechanism influences the similarities and differences observed across multi-layer cases.

In a chain, Kresling units are arranged in series, so that each unit experiences the same force $\tilde{F}_0$. This uniform loading enables us to determine the strain of each unit by applying force equilibrium conditions \citep{https://doi.org/10.1002/advs.202202883}. The equilibrium equation for the $n$-layer chain is given by:
\begin{equation}
    \tilde{F}_0 = \tilde{F}_1 = \tilde{F}_2 = \tilde{F}_3 = \dots = \tilde{F}_n,
\end{equation}
where $\tilde{F}_i$ and $\tilde{F}_0$ refer to the normalised reaction force of the $i$th unit cell and the external force applied on the chain, respectively. Through the identical reaction force of each unit cell, the individual strain $\tilde{u}_i$ can be obtained by solving the following equations:
\begin{equation}
 \begin{bmatrix}
    \tilde{F}_1 = f_1(\tilde{u}_1)\\
    \vdots\\
    \tilde{F}_i = f_i(\tilde{u}_i)\\
    \vdots\\
    \tilde{F}_n = f_n(\tilde{u}_n)
 \end{bmatrix},   
\end{equation}
where $f_i$ is the curve of reaction force of the $i$th unit.

\begin{figure}[pos=ht]
 \begin{center}
  \includegraphics[width=0.75\textwidth]{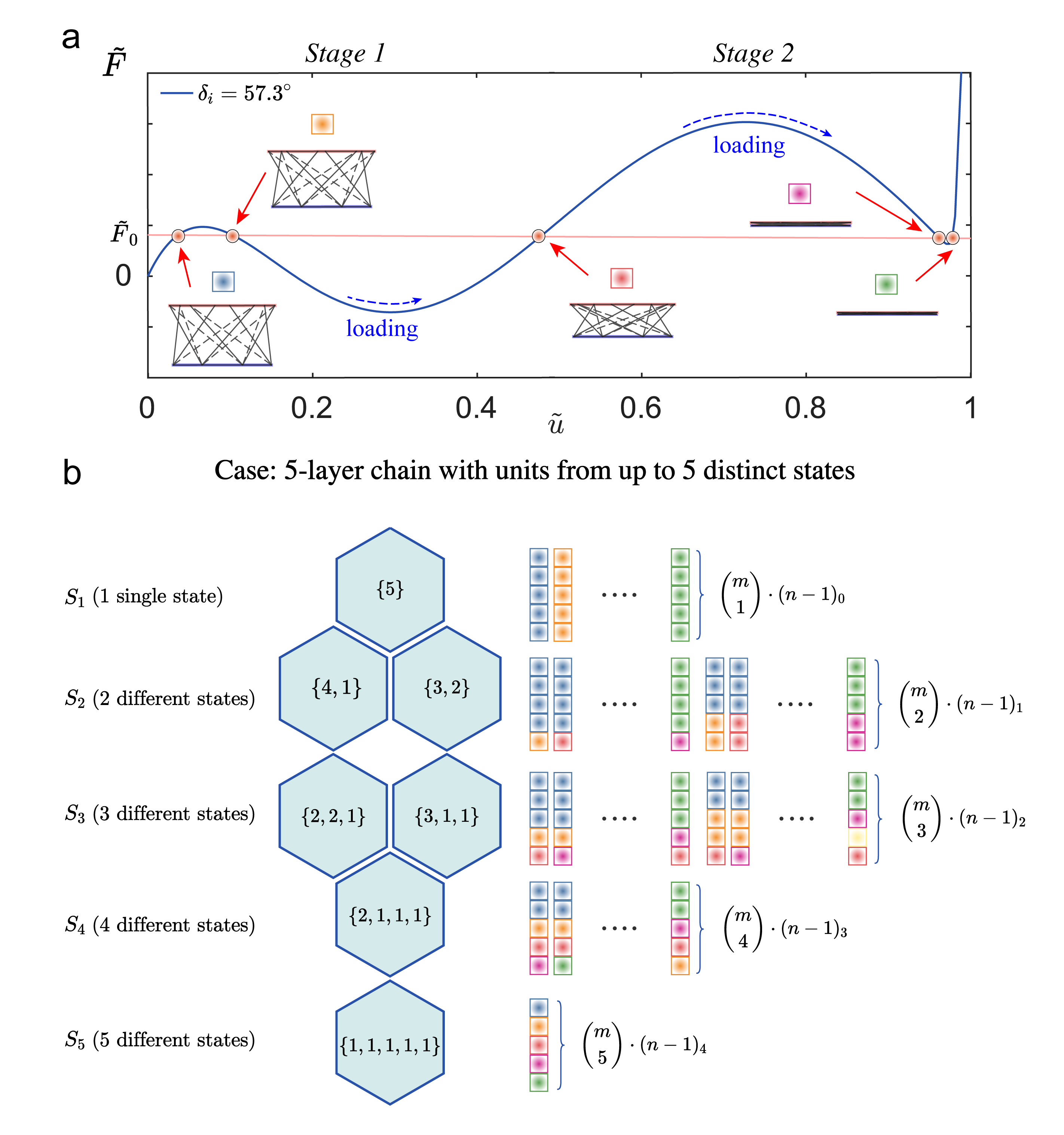}
 \caption{Generalised strategy for solutions of $n$-layer Kresling chain.  (a) Constitutive behaviour of Region III Kresling unit cell, with initial orientation angle $\delta_{i}=57.3^\circ$ and undeformed height-to-radius ${h}^{(0)}/{R^{(0)}}=1$. Under reaction force $\tilde{F}_0$, the unit cells in series can exhibit a maximum of five distinct solutions, denoted by blocks with five distinct colours (blue, yellow, red, pink, green). (b) Example of five-layer Kresling chain comprised of unit cells from a maximum of five distinct states. Combinations are illustrated with coloured blocks, representing the distinct states. The configurations of chain are denoted by permutation expressions on the right. Notation $\{4,1\}$ represents the unordered configuration of chain where four unit cells are in one state, while the remaining one is in another, and so on analogously for other situations with similar notations.}
\label{fig:Homogenization}
 \end{center}
\end{figure}

As demonstrated in \autoref{fig:Homogenization}~(a), a single bistable compact unit cell ($51.0^\circ<\delta_{i}<67.8^\circ$) can exhibit up to five distinct geometric states that coexist under the same externally applied force $\tilde{F}_0$. Consequently, in an $n$-layer chain assembled from such unit cells, the chain’s configuration can be grouped into at most five distinct unit-cell states, each characterised by identical values of $h_i$ and $\gamma_i$. This grouping greatly simplifies the analysis, as the $n$-layer system can be reduced to a 12-variable algebraic system consisting of five pairs of $(h_i,\gamma_i)$, along with $h$ and $F$.

Having established this reduced representation, we now incorporate it into a solution procedure to simplify the computation of equilibrium paths. This approach is exemplified using the five-layer system in \autoref{fig:Homogenization}~(b), which can accommodate up to five distinct unit-cell states simultaneously. To describe how each state is distributed across the chain, we introduce the following notation. If all five units are in the same state ($S_1$), the chain configuration is denoted as $\{5\}$. For two distinct states ($S_2$), the configurations could be $\{4,1\}$ or $\{3,2\}$, where the numbers indicate the counts of units in each state, for instance, $\{4,1\}$ means four units are in one state and one unit is in the other. Similarly, a chain with three distinct states ($S_3$) can be represented as $\{3,1,1\}$ or $\{2,2,1\}$, corresponding to different ways of partitioning the five units among three states. The notations for cases with four ($S_4$) and five ($S_5$) follows the same logic.

Since the unit cells under the same state exhibit identical kinematic responses across both $h_i$ and $\gamma_i$, additional constraints for the governing algebraic system can be derived as:
\begin{equation}
\left \{
\begin{aligned}
H_{\mathrm{constraint}} &= h_{i} - h_{j}=0 \\
\Gamma_{\mathrm{constraint}} &=\gamma_{i} - \gamma_{j}=0 
\end{aligned}
\right .
\quad \forall i, j \in G_{l}, \quad l \in {1,2,...,5},
\label{eq:Force equilibrium}
\end{equation}
where $G_l$ refers to the group of unit cells under identical deformation state.

\begin{figure}
 \begin{center}
  \includegraphics[width=0.95\textwidth]{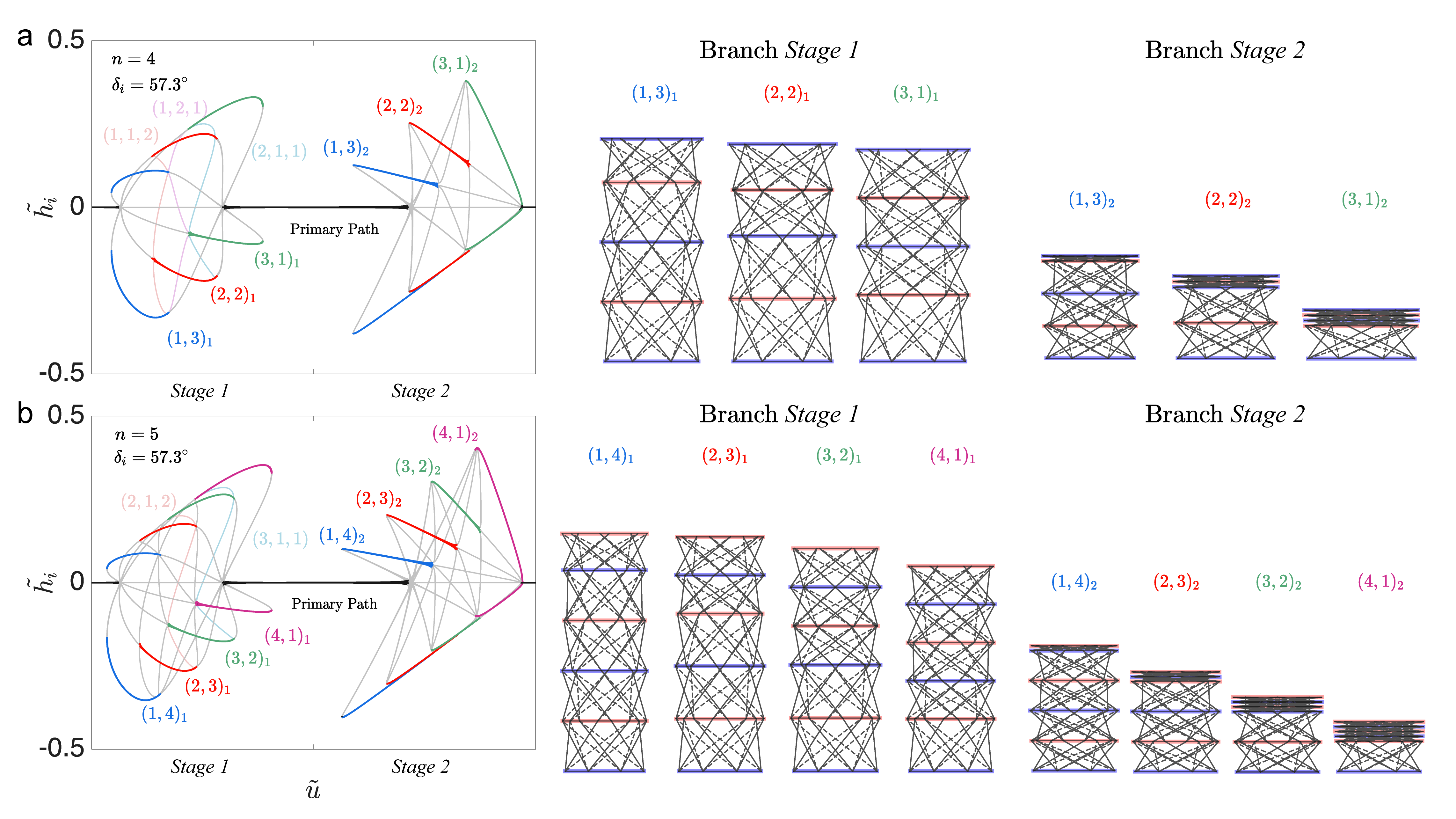}
 \caption{Bifurcation diagram of the $n$-layer Kresling chain assembled with identical unit cells from Region III ($\delta_{i}=57.3^\circ$). Bold and light lines refer to stable and unstable equilibrium. Odd-layered case ($n=4$) and even-layered case ($n=5$) are compared. (a)-(b) Bifurcation diagram described by individual height ${\tilde{h}_{i}}$ with respect to strain $\tilde{u}$. Notation $(1,3)_1$, $(2,2)_2$, etc., refer to ordered configurations of branches in both bifurcation Stage 1 and 2. Note that $(1,3)$ explicitly represents configurations with orders, distinguishing between $(1,3)$ and $(3,1)$ within $\{3,1\}$, as illustrated in \autoref{fig:Homogenization}. Corresponding deformed configurations are demonstrated on the right side.}
\label{fig:n layer bifurcation diagram}
 \end{center}
\end{figure}

\subsection{Predictable geometric loci for multi-stage bifurcation}
\label{sec:Predictable geometric loci for multi-stage bifurcation}

\begin{figure}[pos=p]
 \begin{center}
  \includegraphics[width=0.9\textwidth]{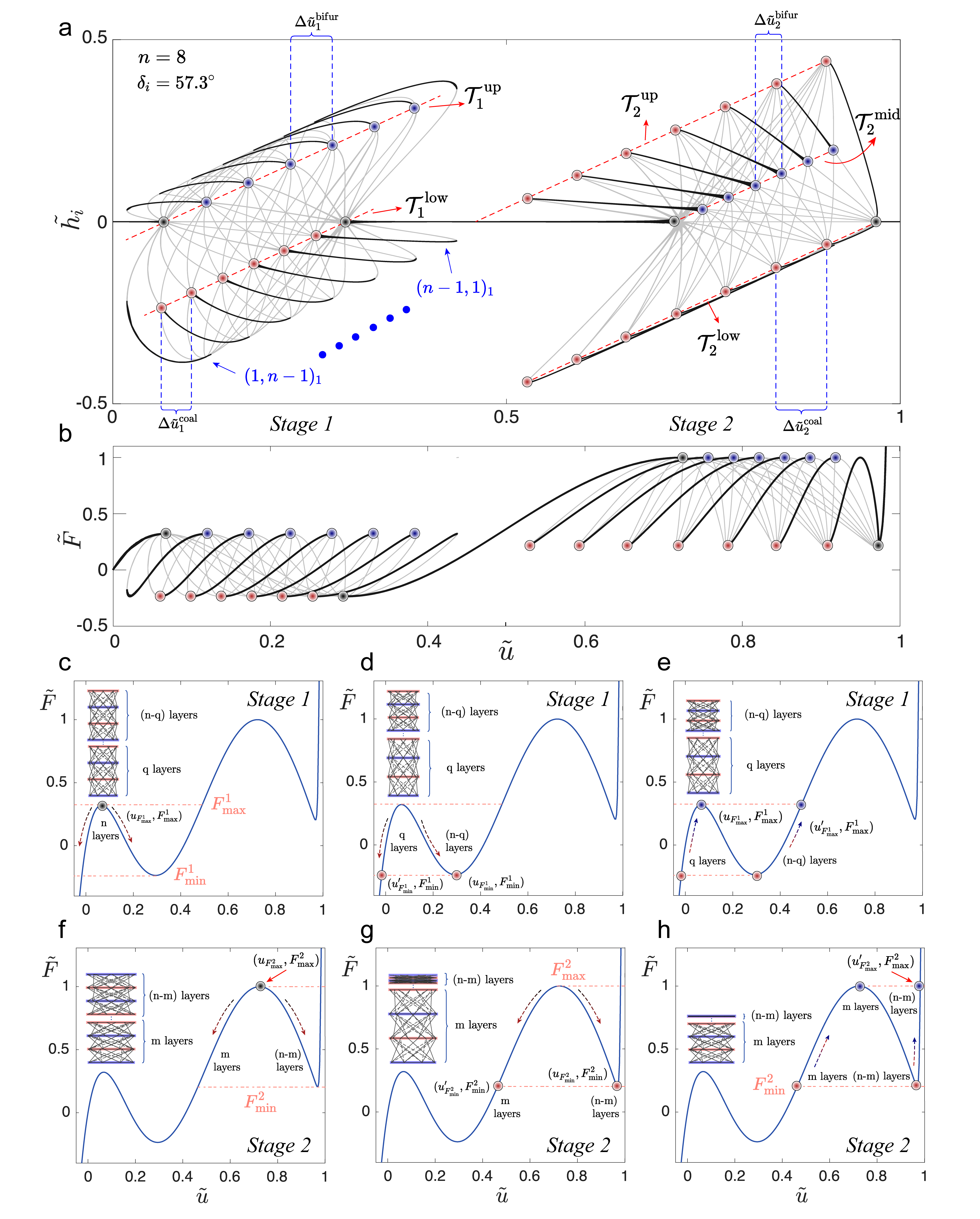}
 \caption{Bifurcation of $n$-layer Kresling chain assembled with identical unit cells from Region III ($\delta_{i}=57.3^\circ$), taking $n=8$ as an example. The primary (black), secondary (blue) bifurcation points, and secondary coalescence (red) points are demonstrated. (a)-(b) Bifurcation diagram and curve of reaction force $\tilde{F}$. Bold and light lines refer to stable and unstable equilibrium. Trajectories in bifurcation Stage 1 ($\mathcal{T}_{1}^{\mathrm{low}}$,$\mathcal{T}_{1}^{\mathrm{up}}$), Stage 2 ($\mathcal{T}_2^{\mathrm{low}}$,$\mathcal{T}_2^{\mathrm{mid}}$,$\mathcal{T}_2^{\mathrm{up}}$), and spacing of bifurcation points $\Delta \tilde{u}_1^{\mathrm{bifur}}$, $\Delta \tilde{u}_1^{\mathrm{coal}}$, $\Delta \tilde{u}_2^{\mathrm{bifur}}$, and $\Delta \tilde{u}_2^{\mathrm{coal}}$ are illustrated. (c)-(e) Reaction force $\tilde{F}$ and deformed configurations of corresponding unit cell in Stage 1. (f)-(h) Reaction force $\tilde{F}$ and deformed configurations of corresponding unit cell in Stage 2.}
\label{fig:n-layer, unit cell correspondence} 
 \end{center}
\end{figure}

The introduction of these constraints into the governing algebraic system enables the construction of bifurcation diagrams for any $n$-layer homogeneous chain. To illustrate, we present bifurcation diagrams and corresponding unit cell configurations for the four-layer, five-layer, and eight-layer cases in \autoref{fig:n layer bifurcation diagram} and \autoref{fig:n-layer, unit cell correspondence}, highlighting the similarities in bifurcation as $n$ increases. For chains assembled from Region III unit cells, the bifurcation diagrams consistently exhibit two distinct branching stages (Stage 1 and 2), marked by the sequential evolution through primary, secondary, and tertiary paths, followed by eventual coalescence. Notably, in this section, the notation $(1,3)$ is used to denote an ordered configuration of the chain. This allows the branch $(1,3)$ to be distinguished from $(3,1)$, although both correspond to the same unordered state $\{3,1\}$. The primary path, corresponding to configuration $S_1$, undergoes bifurcation into multiple secondary branches, each corresponding to a configuration $S_2$. For example, in the four-layer case, there are three pairs of secondary branches: $(1,3)$, $(2,2)$, and $(3,1)$. In the 5-layer case, a similar pattern emerges, with four pairs of secondary branches, $(1,4)$, $(2,3)$, $(3,2)$, and $(4,1)$, emanating from the primary path. The number of secondary branches is determined by the possible ways to partition the chain into distinct $S_2$ configurations.

Tertiary branches delineates the stable segments of the secondary paths. Originating from these critical branching points, the tertiary branches serve as interconnecting bridges within the bifurcation landscape. For instance, in Stage 1 of the four-layer case (\autoref{fig:n layer bifurcation diagram}~(a)), although unstable, the $(1,3)$ branch is directly connected to both the $(2,2)$ and $(3,1)$ branches via tertiary branches $(1,1,2)$ and $(1,2,1)$. Similarly, the secondary $(2,2)$ and $(3,1)$ branches are connected through the intermediate tertiary branch $(2,1,1)$. In general, the number of tertiary branches in an $n$-layer system is given by $\binom{n-1}{2}$, representing all possible interconnections between secondary branches. And the number of secondary bifurcation and coalescence points are given by $4(n-2)$ for Stage 1 and $3(n-2)$ for Stage 2. This combinatorial relationship holds as $n$ increases, so the network of tertiary branches expands systematically with the system’s complexity. 

As further confirmation, the five-layer and eight-layer case shown in \autoref{fig:n layer bifurcation diagram}~(b) and \autoref{fig:n-layer, unit cell correspondence}~(a) demonstrates that the observed number of secondary and tertiary branches aligns with this mathematical pattern. This consistent relationship is also observed in bifurcation Stage 2.

As the number of layers in the system increases, so does the number of unstable states, resulting in an increase of sequential snap-through events under displacement control. This is reflected in the reaction force $\tilde{F}$ versus strain $\tilde{u}$ curve for the $n$-layer system (\autoref{fig:n-layer, unit cell correspondence}~(b)), which shows a similar overall trend to the two-layer and three-layer systems, but with more discontinuous stable equilibrium branches. Notably, after each snap-through event, the reloading stiffness along the stable equilibrium paths from secondary coalescence points (red) to bifurcation points (blue) evolves differently depending on the stage: in Stage 1, as more layers collapse, the overall reloading stiffness progressively decreases, while in Stage 2, further collapse of layers leads to an increase in the overall reloading stiffness.

Within Stages 1 and 2, an additional notable feature of the bifurcation emerges: the primary and secondary bifurcation and coalescence points tend to align along five distinct linear trajectories, as illustrated in \autoref{fig:n-layer, unit cell correspondence}~(a) for the eight-layer chain. Here, a “trajectory” refers to a line along which several bifurcation or coalescence points align. In Stage 1, two trajectories are observed. The upper trajectory, $\mathcal{T}_1^{\mathrm{up}}$, is formed by secondary bifurcation points. The lower trajectory, $\mathcal{T}_1^{\mathrm{low}}$, consists of secondary coalescence points. In Stage 2, there are three trajectories: $\mathcal{T}_2^{\mathrm{up}}$ (upper) and $\mathcal{T}_2^{\mathrm{low}}$ (lower), both formed by secondary coalescence points, and a middle trajectory, $\mathcal{T}^{\mathrm{mid}}_2$, formed by primary and secondary bifurcation points. To investigate this pattern, we map the chain’s response onto the constitutive behaviour of the Kresling unit cell (\autoref{fig:n-layer, unit cell correspondence}~(c)-(e), (f)-(h)). For each unit in the $n$-layer chain, we indicate its corresponding compression state on this plot as the chain is loaded. This approach reveals how the deformation and force state of every unit evolves throughout the loading process and illustrates how these individual contributions collectively determine the overall response and bifurcation of the chain. For additional clarity, we provide illustrations below the constitutive behaviour plots that show the geometry and corresponding state of each unit within the $n$-layer chain for Stages 1 and 2.

In Stage 1, the system deforms uniformly along the primary path. This continues until the reaction force reaches its maximum, at $(u_{F^1_{\mathrm{max}}}, {F^1_{\mathrm{max}}})$ (see \autoref{fig:n-layer, unit cell correspondence}(c)). At this point, the chain bifurcates into two distinct states, marking the transition from the $S_1$ to the $S_2$ configuration. To maintain force equilibrium after bifurcation, $q$ out of $n$ layers undergo slight extension, while the remaining $(n-q)$ layers are further compressed (\autoref{fig:n-layer, unit cell correspondence}~(d)). As the reaction force $\tilde{F}$ decreases to $F^1_{\mathrm{min}}$, the chain reaches the secondary coalescence point (red). Here, different values of $q$ correspond to different secondary branches. As the force increases again, these two partitioned states undergo further compression, eventually reaching the secondary bifurcation points (blue), shown in \autoref{fig:n-layer, unit cell correspondence}~(e).

The evolution of behaviour in Stage 2 is similar, along the primary path, all units remain in the same state, with $\tilde{F}$ varying from the minimum force $(u_{F^1_{\mathrm{min}}},F^1_{\mathrm{min}})$, to the maximum force, $(u_{F^2_{\mathrm{max}}},F^2_{\mathrm{max}})$ (\autoref{fig:n-layer, unit cell correspondence}~(f)). Upon reaching $(u_{F^2_{\mathrm{max}}},F^2_{\mathrm{max}})$, the chain bifurcates onto secondary branches, as demonstrated in \autoref{fig:n-layer, unit cell correspondence}~(g). In these branches, $(n-m)$ layers undergo further compression while the remaining $m$ layers experience release. During this process, $\tilde{F}$ decreases from $F^2_{\mathrm{max}}$ to $F^2_{\mathrm{min}}$. Once $F^2_{\mathrm{min}}$ is reached, the $(n-m)$ layers continue to compress from $(u_{F^2_{\mathrm{min}}},F^2_{\mathrm{min}})$; however, to maintain equilibrium, the $m$ layers stop releasing and begin to compress again. As $\tilde{F}$ increases back to $F^2_{\mathrm{max}}$, the equilibrium path arrives at the secondary bifurcation points, as displayed in \autoref{fig:n-layer, unit cell correspondence}~(h). Here, different values of $m$ correspond to distinct combinations of $S_2$ states, referring to different secondary paths, bifurcation and coalescence points (e.g. $n-m=4,m=1  \text{ corresponds to } \text{branch }(4,1)$), as illustrated by \autoref{fig:n-layer, unit cell correspondence}~(a).

Based on the above insights, the analytical expressions can then be confirmed and validated for the trajectories formed by the secondary bifurcation and coalescence points in Stage 1. Derived from the constitutive behaviour of a Kresling unit, the analytical expression can be written as:
\begin{subequations}
\begin{align}
\mathcal{T}_1^{\mathrm{up}}: \tilde{h}_{i} &= \tilde{u} - u_{F^1_{\mathrm{max}}},\\
\mathcal{T}_1^{\mathrm{low}}: \tilde{h}_{i} &= \tilde{u} - u_{F^1_{\mathrm{min}}}.
\end{align}
\end{subequations}
Following a similar procedure, the analytical expressions for the trajectories in Stage 2 can be derived as:
\begin{subequations}
\begin{align}
\mathcal{T}_2^{\mathrm{up}}: \tilde{h}_{i} &= \tilde{u} - u'_{F^2_{\mathrm{min}}},\\
\mathcal{T}_2^{\mathrm{mid}}:  \tilde{h}_{i} &= \tilde{u} - u_{F^2_{\mathrm{max}}},\\
\mathcal{T}_2^{\mathrm{low}}: \tilde{h}_{i} &= \tilde{u} - u_{F^2_{\mathrm{min}}}.
\end{align}
\end{subequations}

Beyond the linear alignment of the trajectories, it is observed that the secondary bifurcation and coalescence points exhibit uniform horizontal spacing across Stage 1 and 2, which can be written as:
\begin{subequations}
\begin{align}
\Delta \tilde{u}_1^{\mathrm{bifur}} &= \frac{u^\prime_{F^1_{\mathrm{max}}}-u_{F^1_{\mathrm{max}}}}{n},\\
\Delta \tilde{u}_1^{\mathrm{coal}} &= \frac{u_{F^1_{\mathrm{min}}}-u^\prime_{F^1_{\mathrm{min}}}}{n},\\
\Delta \tilde{u}_2^{\mathrm{bifur}} &= \frac{u^\prime_{F^2_{\mathrm{max}}}-u_{F^2_{\mathrm{max}}}}{n},\\
\Delta \tilde{u}_2^{\mathrm{coal}} &= \frac{u_{F^2_{\mathrm{min}}}-u^\prime_{F^2_{\mathrm{min}}}}{n}.
\end{align}
\end{subequations}
This constant interval suggests that the snap-through between successive equilibrium states occurs at regular increments of the Kresling chain's global displacement. Furthermore, this regularity implies that the branch points determined without solving the governing equations of the whole chain. Given the prescribed geometry of Kresling units, the positions of the bifurcation points can be identified, eliminating the need of computational resources. It also provides the inspiration and foundation for inverse designing Kresling-inspired architected materials with predictable bifurcation. 

\section{Conclusions}
\label{Conclusion and Discussion}

The presented work investigated the nonlinear and multi-stable mechanics of Kresling chains by tracking equilibrium paths and bifurcation structures in systems composed of two, three, and ultimately $n$-layers. Particular emphasis was placed on establishing relationships between the mechanical response and the geometric design variables defining the system, including polygon count, initial twist angle, height, radius, and crease lengths. The crease lines were modeled as axial-load-carrying elements, and the governing equations were formulated as a constrained algebraic system solved using continuation and bifurcation analysis. 
Stability was evaluated through the positive definiteness of the projected Hessian of the total strain energy. Starting from the two-layer system, four distinct folding and bifurcation modes were identified depending on the initial orientation angle $\delta_i$: (I) stable supercritical bifurcation with remerging behaviour, (II) unstable remerging behaviour, (III) coexistence of supercritical and subcritical bifurcations, and (IV) purely subcritical bifurcation behaviour. Supercritical bifurcations produced stable post-buckling branches, whereas subcritical bifurcations generated unstable branches. A parametric study over the geometric design space established phase diagrams that map these regimes as functions of $\delta_i$ and the normalised height-to-radius ratio revealing progressive stabilisation of the phase boundaries as $N$ increases. The analysis was then extended to three-layer chains by introducing additional twist and axial DOF. Compared with the two-layer case, the three-layer system exhibited additional secondary equilibrium branches associated with symmetry-breaking configurations between unit cells, highlighting the increasing complexity of the post-critical response with chain length.
Finally, a reduced-order generalisation strategy was developed to extend the analysis to $n$-layer chains while overcoming the numerical challenges caused by ill-conditioned Jacobians and the rapid growth in DOFs near bifurcation points. By grouping unit cells sharing identical equilibrium states, the $n$-layer problem was reduced to an equivalent five-state representation. This enabled efficient prediction of equilibrium paths and bifurcation structures for chains with four or more layers. The resulting bifurcation diagrams showed that the number of primary, secondary, and tertiary branches corresponds directly to combinations of the equilibrium states. Furthermore, the trajectories of secondary and tertiary bifurcations in Phase II were derived and validated, revealing regularly spaced branches governed by the local extrema of the unit-cell constitutive response.
Overall, the proposed framework directly links geometric design parameters to programmable post-critical responses in Kresling chains, providing a foundation for the inverse design of multi-layer mechanical meta-structures with tailored equilibrium paths, tuneable multi-stability, and controllable deployment.

\printcredits

\section*{Declaration of competing interest}
The authors declare that they have no known competing financial interests or personal relationships that could have appeared to influence the work reported in this paper.
\section*{Data availability}
Data will be made available on request.
\section*{Acknowledgments}
\label{sec: Acknowledgments}
The authors LdW and MAD thank UKRI for support under the EPSRC Open Fellowship scheme (Project No. EP/W019450/1).
\appendix
\section{Derivation of equilibrium path stability}
\label{sec: appendix B}
Once the bifurcation diagram is obtained, the stable and unstable branches of the equilibrium path can then be identified. The stability of equilibrium path is determined by the definiteness of Hessian matrix derived by the function of total potential energy. The bordered Hessian matrix of total potential energy $E$ is given by:
\begin{equation}
\label{eq:hessian matrix}
\begin{aligned}
&\mathbf{H}_E =
\begin{bmatrix}
\nabla^2_{xx} \mathcal{L} & {\nabla g}^{T}\\
\nabla g & 0
\end{bmatrix}
=
\begin{bmatrix}
\frac{\partial^2 E}{\partial h_1^2} & \frac{\partial^2 E}{\partial h_1 \partial \gamma_1} &  \cdots & \frac{\partial^2 E}{\partial h_1 \partial h_n} & \frac{\partial^2 E}{\partial h_1 \partial \gamma_n} & \frac{\partial^2 E}{\partial h_1 \partial F}\\
\frac{\partial^2 E}{\partial \gamma_1 \partial h_1} & \frac{\partial^2 E}{\partial \gamma_1^2} &  \cdots & \frac{\partial^2 E}{\partial \gamma_1 \partial h_n} & \frac{\partial^2 E}{\partial \gamma_1 \partial \gamma_n} & \frac{\partial^2 E}{\partial \gamma_1 \partial F}\\
\vdots & \vdots & \ddots & \vdots & \vdots & \vdots \\
\frac{\partial^2 E}{\partial h_n \partial h_1} & \frac{\partial^2 E}{\partial h_n \partial \gamma_1} &  \cdots & \frac{\partial^2 E}{\partial h_n^2} & \frac{\partial^2 E}{\partial h_n \partial \gamma_n} & \frac{\partial^2 E}{\partial h_n \partial F}\\
\frac{\partial^2 E}{\partial \gamma_n \partial h_1} & \frac{\partial^2 E}{\partial \gamma_n \partial \gamma_1} &  \cdots & \frac{\partial^2 E}{\partial \gamma_n \partial h_n} & \frac{\partial^2 E}{ \partial \gamma_n^2} & \frac{\partial^2 E}{\partial \gamma_n \partial F}\\
\frac{\partial^2 E}{\partial F \partial h_1} & \frac{\partial^2 E}{\partial F \partial \gamma_1} &  \cdots & \frac{\partial^2 E}{\partial F h_n} & \frac{\partial^2 E}{\partial F \partial \gamma_n} & \frac{\partial^2 E}{\partial F^2}
\end{bmatrix},\\ 
&\text{s.t.} \quad g(h,h_{i}) = h - \sum_{i=1}^n h_{i}
=0,
\end{aligned}
\end{equation}
where, $\nabla^2_{xx}\mathcal{L}$ refers to the unbordered Hessian matrix.

Derived from Eq.\eqref{eq:EQUILIBRIUM CRITERION}, only four types of elements in the Hessian matrix are none-zero, which can be written as:
\begin{subequations}
\begin{align}
\frac{\partial^2 E}{\partial {h_{i}}^2} = &N\left(\mathrm{k}_{\mathrm{v}}\left(1-\frac{L^{(0)}_{\mathrm{v}_i}}{L_{\mathrm{v}_i}}\right)+\mathrm{k}_{\mathrm{d}}\left(1-\frac{L^{(0)}_{\mathrm{d}_i}}{L_{\mathrm{d}_i}}\right)\right)+{h_{i}}^2N\left(\mathrm{k}_{\mathrm{v}}\frac{L^{(0)}_{\mathrm{v}_i}}{{L_{\mathrm{v}_i}}^3}+\mathrm{k}_{\mathrm{d}}\frac{L^{(0)}_{\mathrm{d}_i}}{{L_{\mathrm{d}_i}}^3}\right )+\frac{P} {\epsilon} \frac{e^{-h_{i}/\epsilon}}{\left(1+e^{-h_{i}/\epsilon}\right)^2},\\
\begin{split}
\frac{\partial^2 E}{\partial {\gamma_{i}}^2} = &N{R^{(0)}}^2\left (\mathrm{k}_{\mathrm{v}}\left(1-\frac{L^{(0)}_{\mathrm{v}_i}}{L_{\mathrm{v}_i}}\right)\cos(\delta_{i}+\gamma_{i})+\mathrm{k}_{\mathrm{d}}\left(1-\frac{L^{(0)}_{\mathrm{d}_i}}{L_{\mathrm{d}_i}}\right)\cos\left(\delta_{i}+\gamma_{i}+\frac{2\pi}{N}\right)\right)\\
+&N{R^{(0)}}^4\left (\mathrm{k}_{\mathrm{v}}\sin^2(\delta_{i}+\gamma_{i})\frac{L^{(0)}_{\mathrm{v}_i}}{{L_{\mathrm{v}_i}}^3}+\mathrm{k}_{\mathrm{d}}\sin^2\left(\delta_{i}+\gamma_{i}+\frac{2\pi}{N}\right)\frac{L^{(0)}_{\mathrm{d}_i}}{{L_{\mathrm{d}_i}}^3}\right )
\end{split},\\
\frac{\partial^2 E}{\partial h_{i} \partial \gamma_{i}} = & h_{i}N{R^{(0)}}^2\left (\mathrm{k}_{\mathrm{v}}\frac{L^{(0)}_{\mathrm{v}_i}}{{L_{\mathrm{v}_i}}^3}\sin(\delta_{i}+\gamma_{i})+\mathrm{k}_{\mathrm{d}}\frac{L^{(0)}_{\mathrm{d}_i}}{{L_{\mathrm{d}_i}}^3}\sin\left(\delta_{i}+\gamma_{i}+\frac{2\pi}{N}\right)\right ),\\
\frac{\partial^2 E}{\partial h_{i} \partial F} = &-1.
\end{align}
\end{subequations}
The definiteness of bordered Hessian matrix determines the stability of the potential energy function $E$. However, a significant numerical challenge arises near the bifurcation point: the bordered Hessian matrix tends to be singular, which makes it harder to evaluate the stability of the equilibrium path. Then the second-order conditions described by bordered Hessian matrix $H_E$ can be stated in the form of projected Hessian matrix $H_p$ onto subspaces with the given constraints $g(h,h_{i})$. The projected Hessian matrix is then given by \citep{Nocedal2006}:
\begin{equation}
H_p = Z^T\nabla^2_{xx}\mathcal{L}Z,
\end{equation}
where, $Z$ is defined as the basis whose columns spans the null space of $\nabla g$:
\begin{equation}
    Z = \text{Null}(\nabla g).
\end{equation}

By evaluating the definiteness of the projected Hessian matrix, the stability of branches can be determined, as demonstrated in this article.

\section{Governing algebraic system with extra constraints}
\label{sec: appendix C}

The total potential energy of $n$-layer Kresling chain with penalty term is given by:
\begin{equation}
E = \sum_{i=1}^{n} N\left( \frac{1}{2} \mbox{k}_{\mathrm{v}}  {\Delta s_{\mathrm{v}_i}}^2 + \frac{1}{2} \mbox{k}_{\mathrm{d}}  {\Delta s_{\mathrm{d}_i}}^2\right )+F (h-\sum_{i=1}^{N}h_{i}) + {P}\,\lim_{\epsilon\rightarrow0} \sum_{i=1}^{n}  \,\epsilon\,\ln\left(1+e^{-h_{i}/\epsilon}\right).
\label{eq:algebraic without constraints}
\end{equation}
By adding the extra constraints Eq.\eqref{eq:Force equilibrium} to Eq.\eqref{eq:algebraic without constraints}, the total potential energy can be written as:
\begin{equation}
E = \sum_{l=1}^{S}b_lN\left ( \frac{1}{2} \mbox{k}_{\mathrm{v}}  {\Delta s_{\mathrm{v}_l}}^2 + \frac{1}{2} \mbox{k}_{\mathrm{d}}  {\Delta s_{\mathrm{d}_l}}^2\right ) + F (h-\sum_{l=1}^{S}b_l h_{l}) + P\,\sum_{l=1}^{S}b_l\, \epsilon\,\ln\left(1+e^{- {h_{l}}/{\epsilon}}\right),
\end{equation}
 where $S$ corresponds to the total number of the possible deformation state of single Kresling unit under load $\tilde{F}_0$ (e.g., for $S_k$ configuration, $S=k$), and $b_l$ denotes the number of individual layers possessing identical deformation (\emph{i.e.} the number of layers within $l$th group $G_l$ indicated by Eq.\eqref{eq:Force equilibrium}). For example, where the maximum number of deformed states $S$ is five, this means that the maximum number of equations within the algebraic system is $2S+1=11$. The algebraic system can then be simplified as:
\begin{subequations}
\begin{align}
&\frac{\partial E}{\partial h_{l}}=N  b_lh_{l}\left\{\mbox{k}_{\mathrm{v}}\left(1-\frac{L^{(0)}_{\mathrm{v}_l}}{L_{\mathrm{v}_l}}\right)+\mbox{k}_{\mathrm{d}}\left(1-\frac{L^{(0)}_{\mathrm{d}_l}}{L_{\mathrm{d}_l}}\right)\right\}+{P} b_l \left(\frac{e^{- h_{l}/\epsilon}}{1+e^{- h_{l}/\epsilon}}\right)-F b_l,\\
&\frac{\partial E}{\partial \gamma_{l}}=N  b_l\left \{\mbox{k}_{\mathrm{v}}\left(1-\frac{L^{(0)}_{\mathrm{v}_l}}{L_{\mathrm{v}_l}}\right)\sin(\delta_{l}+\gamma_{l})+\mbox{k}_{\mathrm{d}}\left(1-\frac{L^{(0)}_{\mathrm{d}_l}}{L_{\mathrm{d}_l}}\right)\sin\left(\delta_{l}+\gamma_{l}+\frac{2\pi}{N}\right)\right \}, \\
&\frac{\partial E}{\partial F}=h-\sum^{S}_{l=1}b_l h_{l}.
\end{align}
\label{eq:simplified algebraic}
\end{subequations}
By achieving this, the algebraic system shown in Eq.\eqref{eq:simplified algebraic} can represent any n-layer Kresling chain system. By solving these equations with AUTO 07P, the bifurcation diagrams of $n$-layer chain can be obtained, as exemplified by the figures in \autoref{sec:Homogenization solution} \citep{doedel2007auto}. 

\bibliographystyle{model1-num-names}

\bibliography{Refs}



\end{document}